\documentclass[10pt, journal, comsoc]{IEEEtran}

\usepackage{cite}
\usepackage{amsmath,amssymb,amsfonts}
\usepackage{algorithm}
\usepackage{algorithmic}
\usepackage{graphicx}
\usepackage{textcomp}
\usepackage{subfigure}
\usepackage{setspace}
\usepackage{longtable}
\usepackage{multirow}
\usepackage{dsfont}
\usepackage{enumerate}
\usepackage{color}
\usepackage{booktabs}
\usepackage{blindtext}

\allowdisplaybreaks[2]
\newtheorem{theorem}{Theorem}

\newtheorem{remark}{Remark}
\newcommand{\IND}{~~~}

\begin{document}

\title{
Joint Switch-Controller Association and Control Devolution
for SDN Systems: An Integration of Online Control and Online Learning
}

\author{
Xi~Huang,~\IEEEmembership{Student Member,~IEEE}, 
Yinxu~Tang,
Ziyu~Shao$^*$,~\IEEEmembership{Senior Member,~IEEE}, \\
Yang~Yang,~\IEEEmembership{Fellow,~IEEE},
Hong~Xu,~\IEEEmembership{Senior Member,~IEEE}
\thanks{
{$^*$ The corresponding author of this work is Ziyu Shao.
		
	\indent X. Huang, Y. Tang, Z. Shao, and Y. Yang are with the School of Information Science and Technology, ShanghaiTech University, China. (E-mail:\{huangxi, tangyx, shaozy, yangyang\}@shanghaitech.edu.cn)
	
	\indent	H. Xu is with Department of Computer Science, City University of Hong Kong, Hong Kong. (E-mail: henry.xu@cityu.edu.hk)}
}
}

\maketitle

\begin{abstract} 
	In software-defined networking (SDN) systems, it is a common practice to adopt a multi-controller design and control devolution techniques to improve the performance of the control plane.
	However, in such systems the decision-making for joint switch-controller association and control devolution often involves various uncertainties, \textit{e.g.}, the temporal variations of controller accessibility, and computation and communication costs of switches. 
	In practice, statistics of such uncertainties are unattainable and need to be learned in an online fashion, calling for an integrated design of learning and control. 
	In this paper, we formulate a stochastic network optimization problem that aims to minimize time-average system costs and ensure queue stability. 
	By transforming the problem into a combinatorial multi-armed bandit problem with long-term stability constraints, we adopt bandit learning methods and optimal control techniques to handle the exploration-exploitation tradeoff and long-term stability constraints, respectively. 
	Through an integrated design of online learning and online control, we propose an effective Learning-Aided Switch-Controller Association and Control Devolution (\textit{LASAC}) scheme. 
	Our theoretical analysis and simulation results show that LASAC achieves a tunable tradeoff between queue stability and system cost reduction with a sublinear time-averaged regret bound over a finite time horizon.
\end{abstract}


\IEEEpeerreviewmaketitle

\section{Introduction}\label{sec:introduction}
\IEEEPARstart{O}{ver} the past decade, software-defined networking (SDN) has emerged to facilitate more efficient network management and more flexible network programmability \cite{kreutz2015software}.
The key idea of SDN is to separate the control plane from the data plane. In this way, the control plane maintains a logically centralized view to orchestrate the whole network, leaving the data plane to carry out basic network functions such as monitoring and packet forwarding.
In the data plane, the forwarding devices such as SDN-enabled switches\cite{gushchin2015scalable} constantly generate requests to be processed by the control plane, \textit{e.g.}, to make routing decisions for dedicated flows. 

As data plane expands, the control plane may become a bottleneck in SDN systems for scalability and reliability concerns. 
Existing works adopt two approaches to control plane design. 
One approach is to implement the control plane with physically distributed controllers. 
In such a design, each switch may have potential connections to multiple controllers.
Due to temporal variations of system dynamics, each switch's accessible controller set may change over time.
Accordingly, the key challenge in such design lies in how switches should forward requests amongst their associated controllers, that is \textit{switch-controller association}. 
To this end, some recent works have proposed various solutions \cite{koponen2010onix, tootoonchian2010hyperflow, dixit2013towards, krishnamurthy2014pratyaastha, levin2012logically, wang2016dynamic, wang2017efficient, filali2018sdn, lyu2018multi, nguyen2019saco}.
The other approach is to delegate part of request processing that requires only local information onto SDN-enabled switches or controllers near the data plane\cite{curtis2011devoflow, hassas2012kandoo, zheng2015lazyctrl, katta2016cacheflow}, that is \textit{control devolution}, to mitigate control plane's workload.
The key challenge in such design lies in \textit{when} and \textit{whether} switches should, if possible, process requests locally or upload them to the control plane.

The above two approaches are orthogonal to each other and can be further combined.
Nonetheless, such a joint design often involves two concerns.
One concern is about the \textit{tradeoff} among the communication costs (\textit{e.g.}, bandwidth or round-trip time) incurred by uploading requests to the control plane, local computational costs on switches, and queue stability in SDN systems\cite{huang2017dynamic}. 
The other concern comes from various \textit{uncertainties} in SDN systems.
For example, statistics of request traffic and controller accessibility are often not attainable in practice; 
besides, switches' communication costs incurred or computation costs would be revealed only when the corresponding transmission or processing is completed.
Faced with such uncertainties, 
we need an \textit{integrated} design of online learning and online control. 
However, there are several challenges. 
\textit{First}, the learning procedure must deal with the \textit{exploration-exploitation} tradeoff. This is because 
i) over-exploration, \textit{i.e.}, switches actively forward requests to different controllers, may result in not only balanced and stable queue backlogs but also excessive communication cost; 
ii) over-exploitation, \textit{i.e.}, switches stick to sending requests to some controllers with empirically lowest costs, may overload such controllers and miss potentially better candidates.
\textit{Second}, with online learning and online control procedures being coupled, their interplay demands a careful design, because 
1) the learning procedure, if conducted ineffectively, may misguide the control procedure and lead to excessive costs or overloaded queue backlogs; 
2) meanwhile, the control procedure, if carried out improperly, may incur noisy feedback and thus impose adverse effects on learning efficiency.
\textit{Third}, due to various uncertainties, online control without exact prior information would inevitably incur performance loss (\textit{a.k.a.} \textit{regret}). Quantifying such regrets can provide system designers with a better understanding of their design space.

In this paper, we address the above challenges by adopting an effective combination of Lyapunov optimization\cite{neely2010stochastic} and bandit learning\cite{auer2002finite}. More specifically, we make the following contributions.
\begin{enumerate}
	\item[$\diamond$] \textbf{Modeling and Formulation:} 
	We formulate the problem of joint switch-controller association and control devolution with unknown system dynamics as a combinatorial multi-armed bandit (CMAB) problem with long-term queue stability constraints. 
	For online control, we aim to reduce the time-average communication costs and computation and ensure the long-term stability of all queue backlogs.
	For online learning, we aim to minimize regret due to decision making under uncertainty. 
	\item[$\diamond$] \textbf{Algorithm Design:} 
	By exploiting the unique problem structure,
	we devise an effective Learning-Aided Switch-Controller Association and Control Devolution (\textit{LASAC}) scheme, which achieves a proper integration of online control and online learning.  
	Specifically, regarding online control, we adopt optimal control techniques \cite{neely2010stochastic} to handle the long-term queue stability constraints and conduct efficient online decision-making to minimize the total system costs. 
	To achieve effective online learning, we employ \textit{UCB1-tuned}\cite{auer2002finite}, an improved version of classic upper-confidence bound (UCB1) method, to balance the exploration-exploitation tradeoff. 
	\item[$\diamond$] \textbf{Performance Analysis:} Our theoretical analysis shows that LASAC can perform effective online control by achieving an $[O(V), O(1/V)]$ tradeoff between queue stability and system cost reduction with a tunable positive parameter $V$, 
	and realize efficient learning by limiting the time-averaged regret within a sublinear $O(\sqrt{\log{T}/T})$ bound over a finite time horizon $T$. 
	\item[$\diamond$] \textbf{Experimental Verification and Evaluation:} 
		We conduct extensive simulations to evaluate LASAC and its variants. 
		Results from our simulations not only verify the theoretical tradeoffs of LASAC but also show that LASAC and its variants  effectively reduce total system costs with mild-value of regrets and queue stability guarantee.
\end{enumerate}
The rest of this paper is organized as follows. 
We discuss related works in Section \ref{sec: related work}. Then we present our system model and problem formulation in Section \ref{sec: problem formulation}. 
In Section \ref{sec: algorithm design}, we demonstrate the design of LASAC, followed by its performance analysis. 
Next, we analyze our simulation results in Section \ref{sec: simulation}. Finally, we conclude this paper in Section \ref{sec: conclusion}.

\section{Related Work}  \label{sec: related work}
\textbf{Optimizations for SDN Systems}:
So far, a number of existing works \cite{das2019survey} have been conducted to optimize the performance (\textit{e.g.}, the control latency\cite{heller2012controller,zhang2016role,he2017modeling}, resiliency\cite{tanha2016enduring,killi2017capacitated,killi2018placement}, cost efficiency\cite{tanha2016enduring}, \textit{etc.}) of SDN systems. 
Among such works, there are two lines of research in recent years that focus on optimizing the effective control and optimization for switch-controller association \cite{koponen2010onix, tootoonchian2010hyperflow, dixit2013towards, krishnamurthy2014pratyaastha, levin2012logically, wang2016dynamic, wang2017efficient, filali2018sdn, lyu2018multi, nguyen2019saco} and control devolution \cite{curtis2011devoflow, hassas2012kandoo, zheng2015lazyctrl, katta2016cacheflow}, respectively. Instead of studying these two topics separately, later works \cite{huang2017dynamic}\cite{huang2019iwqos} considered the problem of joint switch-controller association and control devolution, then proposed \textit{online} control and predictive control schemes to optimize system costs with queue stability guarantee, respectively.
Although the problem studied by such works is similar to our work, we would like to point out that all such works implicitly assume the \textit{full} availability of instant system dynamics, which is hard to attain in practice. 
In contrast, our work assumes only \textit{partial} availability of instant system dynamics (\textit{e.g.}, queue backlog sizes of controllers but not system costs) upon decision making. 
Generalizing to such more settings brings great challenges and complexities to the scheme design, which are fully addressed by this paper.

\textbf{Learning-aided Control for SDN Systems}:
During the past decade, there has been a growing interest in leveraging machine learning techniques to characterize and improve performances of SDN systems \cite{xie2019survey} such as those for routing optimization\cite{sendra2017including}\cite{lin2016qos}, traffic classification\cite{amaral2016machine}\cite{wang2016framework},
and resource management\cite{chen2017caching,d2017game,he2017algorithm}.
Although the effectiveness of such works have been well justified, most of them were conducted based on \textit{offline} learning techniques with readily available datasets, while \textit{online} learning techniques have been considered by only few works. For example, Rehmani \textit{et al.} \cite{rehmani2018achieving} focused on SDN-based smart grids and proposed an approach based on $\epsilon$-greedy method to learn link failure statistics and direct controllers to adapt switches' coordination to dynamic network conditions. 
Different from such works, our work studies the problem of joint switch-controller association and control devolution with unknown system dynamics in SDN systems. Accordingly, we not only devise an effective scheme with performance guarantee but also characterize the interplay between control and learning. To our best knowledge, it is the first systematic study on the joint control and learning for SDN systems.

\textbf{Learning-aided Control under Bandit Settings}: 
To date, the multi-armed bandit (MAB) model \cite{lattimore2018bandit} has been extensively adopted to study sequential decision-making problems under uncertainty within a wide range of scenarios\cite{bouneffouf2019survey}, such as the cache placement in wireless caching systems \cite{xin2020icc}\cite{junge2020icassp}, the channel allocation in fog systems\cite{junge2020icc},  collaborative filtering in recommendation systems\cite{li2016collaborative}, and beam tracking in millimeter-wave communication systems\cite{aykin2020mamba}. 
Among varieties of MAB settings, the most related to our work is the combinatorial multi-armed bandit model in \cite{li2019combinatorial}. In their model, each arm is assigned with a unique virtual queue to trace the proportion of time being selected, which serves as an indicator to ensure fairness constraints. Each queue backlog is only affected by the selection of its associated arm. 
Different from their setting, in our model, each queue backlog is not uniquely assigned to a particular arm. Instead, the change of queue backlogs during each round may result from the selection of multiple arms. 
Therefore, our model is more complicated and the resulting problem is more challenging to solve. Faced with such difficulties, we not only devise an effective scheme that solves the problem with a sublinear time-averaged regret bound but also characterize the interplay between online control and online learning. Our investigations provide interesting insights for the study of learning-aided approaches.

\begin{figure}
  \centering 
  \includegraphics[scale=.22]{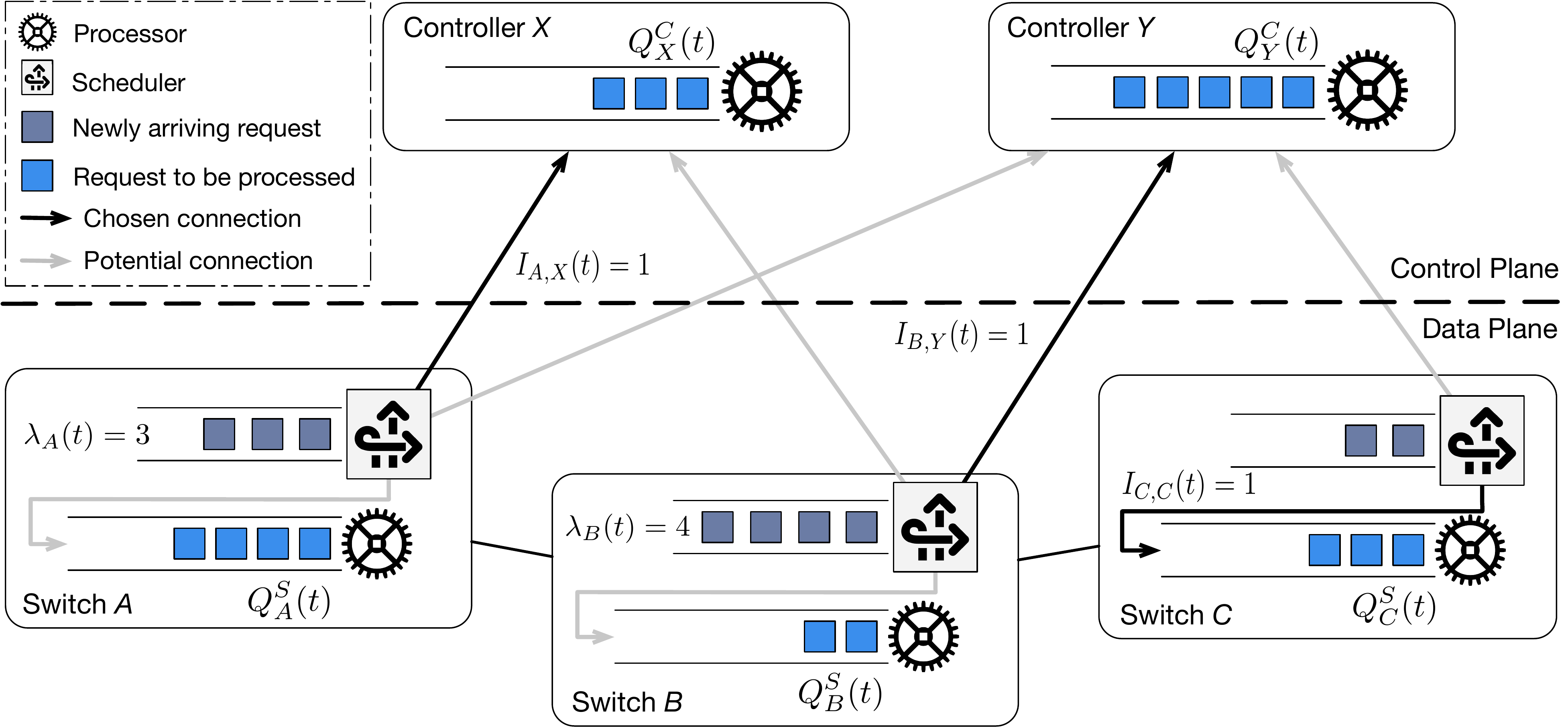}
\caption{
	An instance of our system model in time slot $t$, with three switches in the data plane and two controllers in the control plane. Each switch $i$ generates a number of $\lambda_{i}(t)$ new requests and the scheduler makes decision $I_{i, \cdot}(t)$ to keep the new requests processed locally (\textit{e.g.}, $I_{C, C}(t) = 1$), or forward them to one of its potentially associated controllers (\textit{e.g.}, $I_{A, X}(t) = 1$).
	New requests are then appended to the corresponding queue backlogs and processed in a first-in-first-out manner.}
\label{figure: sys-model}
\end{figure}
\setlength{\textfloatsep}{0pt}

\section{Problem Formulation}  \label{sec: problem formulation}
\subsection{Basic Model}  \label{subsec: overall model}
We consider an SDN system that operates over a finite number of $T$ time slots, as demonstrated in Figure \ref{figure: sys-model}. Each time slot is indexed by $t \in \{0, 1, 2, \dots, T-1\}$.
The control plane and data plane of the system are formed by a set of controllers (denoted by set $\mathcal{C}$) and a set of switches (denoted by set ${\mathcal{S}}$), respectively.
Each switch $i \in \mathcal{S}$ has potential associations to a subset $\mathcal{C}_{i} \subseteq \mathcal{C}$ of controllers.

Due to the temporal variation of SDN system dynamics, the set of accessible controllers for each switch $i$ may change over time. 
We use $A_{i}(t)$ to denote the accessible controller set of switch $i$ during time slot $t$.
By defining set $\mathcal{P}(\mathcal{C}_{i})$ as the power set of $\mathcal{C}_{i}$, we assume that $p(Z_{i}) = \text{Pr}\{{A}_{i}(t) = Z_{i}\}$ for each $Z_{i} \in \mathcal{P}(\mathcal{C}_{i})$. Moreover, each ${A}_{i}(t)$ is assumed \textit{i.i.d.} across time slots and independent across switches.

Next, for all requests in the system, we assume that they are homogeneous and each request can be \textit{processed locally} or \textit{uploaded} to its switch's potentially associated controllers.\footnote{In real systems, however, some requests may be heterogeneous and cannot be processed locally on switches due to limited information and functionality on switches or relevant statefulness on controllers. 
For such cases, our setting can be extended by imposing extra constraints on the scheduling decisions.}

\subsection{System Workflow}
The overall workflow of the system proceeds as follows.
At the beginning of each time slot $t$, 
each switch $i$ generates a number $\lambda_{i}(t)$ of new requests, \textit{e.g.}, \textit{flow install request}\cite{kreutz2015software}. 
The number $\lambda_{i}(t)$ is assumed to be upper bounded by some constant $\lambda_{\text{max}}$.
Then switch $i$ should make its \textit{devolution decision} about whether to process such new requests locally or not. 
If deciding to upload new requests to its potentially associated controllers, then switch $i$ should make its \textit{association decision} about which controller to associate with. 
With decisions being made, some switches upload new requests to controllers while the others process new requests locally. 
Meanwhile, switches and controllers process as many received requests as possible. 
Such a process is repeated during every time slot.

During such a process, owing to recently developed high-speed transmission techniques \cite{ford2017achieving}, 
we assume that the transmission latencies between switches and controllers are much smaller than each time slot's length.
Besides, due to service capacity limits, switches and controllers may not be able to finish all new requests within one time slot. 
To cope with the untreated requests, each switch (or controller) also maintains a processing queue to buffer them.
In the following subsections, we model the scheduling decisions of switches and controllers and their queueing dynamics in detail.

\subsection{Scheduling Decisions}
For each switch $i$, we denote its \textit{devolution decision} by variable $I_{i, i}(t) \in \{0, 1\}$. 
If $I_{i, i}(t) = 0$, then switch $i$ will associate with one of its potentially associated controllers; otherwise, it will append new requests to its own queue and process them locally. 
In the meantime, we denote switch $i$'s \textit{association decision} by variables $\{ I_{i, j}(t) \}_{j \in \mathcal{C}_{i}}$. Specifically, $I_{i, j}(t) = 1$ indicates that controller $j$ is chosen to process new requests and zero otherwise.\footnote{
		In practice, depending on requests' particular states,  
		the decision making can be adapted such that during each time slot, part of new requests get uploaded and others stay locally.}
For simplicity, we use $\textbf{I}_{i}(t) \triangleq \{ I_{i, i}(t) \} \cup \{ I_{i, j}(t) \}_{j \in \mathcal{C}}$ to denote the decisions of switch $i$ in time slot $t$.
We summarize the decisions of all switches by $\textbf{I}(t) \triangleq \{ \mathbf{I}_{i}(t) \}_{i \in \mathcal{S}}$.

Considering the high costs of simultaneous communication with multiple controllers \cite{mantas2019rama}, each switch $i$ is restricted to choose at most one controller in each time slot. Then we have
\begin{equation}\label{constraint: association}
	\begin{array}{c}
		\displaystyle
		I_{i, i}(t) + \sum_{j \in \mathcal{C}_{i}} I_{i, j}(t) = 1,\ \forall\, i \in \mathcal{S}.
	\end{array}
\end{equation}
Besides, recall that $A_{i}(t)$ denotes the set of accessible controllers of switch $i$ during time slot $t$. Then we have 
\begin{equation}\label{constraint: accessibility}
	I_{i, j}(t) = 0,\ \forall j \in \mathcal{C} \backslash {A}_{i}(t).
\end{equation}

\subsection{Queueing Dynamics}
For each switch $i$, we define $Q^{S}_{i}(t)$ as its queue backlog size at the beginning of time slot $t$.  
Correspondingly, for each controller $j$, we define $Q^{C}_{j}(t)$ as its queue backlog size at the beginning of time slot $t$.
Note that due to the homogeneity of requests, each backlog size is equal to the number of requests stored in the queue.

Based on the workflow described in Subsection \ref{subsec: overall model}, the queueing dynamics in the system can be characterized by the following equations. 
For each switch $i \in \mathcal{S}$, 
\begin{equation}\label{queueing dynamics: switches}
	\begin{array}{c}
		Q^{S}_{i}(t+1) = \bigg[ Q^{S}_{i}(t) + I_{i, i}(t) \lambda_{i}(t) - \mu^{S}_{i}(t) \bigg]^{+},
	\end{array}
\end{equation}
where we define operator $[\cdot]^{+} \triangleq \max\{0, \cdot\}$. 
Next, for each controller $j \in \mathcal{C}$, 
\begin{equation}\label{queueing dynamics: controllers}
	\begin{array}{c}
		\displaystyle
		Q^{C}_{j}(t+1) = \bigg[ Q^{C}_{j}(t) + \sum_{i \in \mathcal{S}} I_{i, j}(t) \lambda_{i}(t) - \mu^{C}_{j}(t) \bigg]^{+}.
	\end{array}
\end{equation}

\subsection{Optimization Objectives:} 
On the one hand, the benefit from uploading requests lies in that controllers have adequate service capacities and thus they incur shorter processing latencies than switches. 
Such a benefit may be offset unexpectedly by considerable communication costs (\textit{e.g.}, round-trip time \cite{huang2017dynamic}) of request uploading. 
On the other hand, if most requests are kept processed locally, switches may be overwhelmed by prohibitively high computational costs. 
As a result, the decision-making for joint switch-controller association and control devolution should carefully manage the balance between computation and communication costs. 
Below we specify the definitions of and the constraints on such costs in our model. 

\textbf{Communication Cost:} 
For each switch $i$ and each of its accessible controllers $j \in {A}_{i}(t)$, we denote their communication cost of sending a request during time slot $t$ by $W_{i, j}(t)$ 
($\leq w_{\text{max}}$ for some positive constant $w_{\text{max}}$). 
Such communication costs are assumed \textit{i.i.d.} across time slots with a finite expectation $\mathbb{E}\{ W_{i, j}(t) \} = \bar{w}_{i, j}$. 
Then with decisions $\textbf{I}(t)$, the total communication cost in time slot $t$ is given by
\begin{equation}
	\begin{array}{c}
		\displaystyle
		W(t) \triangleq \hat{W}(\textbf{I}(t)) = 
		\sum_{i \in \mathcal{S}} \sum_{j \in \mathcal{C}} 
		I_{i, j}(t) \lambda_{i}(t) W_{i, j}(t).
	\end{array}
\end{equation}

\textbf{Computational Cost:} For each switch $i$, we denote its computational cost of processing each request locally during time slot $t$ by $M_{i}(t)$ ($\leq m_{\text{max}}$ for some positive constant $m_{\text{max}}$). 
Then its total computational cost in time slot $t$ is given by
\begin{equation}
	\begin{array}{c}
		\displaystyle
		M(t) \triangleq \hat{M}(\textbf{I}(t)) = 
		\sum_{i \in \mathcal{S}} I_{i, i}(t) \lambda_{i}(t) M_{i}(t).
	\end{array}
\end{equation}

\textbf{Queue Stability:} Besides system costs, the stability of queue backlogs in the system is also worth considering. 
Intuitively, achieving queue stability means that no queue backlog on switches or controllers would increase infinitely and eventually lead to system disruption. 
It also ensures that requests would not queue up intensively on any switch or controller, which by \textit{Little's law}\cite{little1961proof} implies that no requests will ever experience excessively long queueing delay. 
In this paper, we define queue stability \cite{neely2010stochastic} as
\begin{equation}\label{constraint: queue stability}
	\limsup_{T \to \infty} \frac{1}{T} \sum_{t=0}^{T-1} \mathbb{E} \bigg\{ 
		\sum_{i \in \mathcal{S}} Q^{S}_{i}(t) + \sum_{j \in \mathcal{C}} Q^{C}_{j}(t)
	\bigg\} < \infty.
\end{equation}

\subsection{Problem Formulation:}
In our model, we aim to minimize the time-average expectation of the total costs of communication and computation over a finite time horizon $T$, subject to long-term queue stability.
Our problem formulation is given as follows.
\begin{equation}\label{problem formulation: 1}
	\begin{array}{cl}
		\underset{ \{\textbf{I}(t)\}_{t=0}^{T-1} }{\text{Minimize}} & \displaystyle
		\frac{1}{T} \sum_{t=0}^{T-1} \mathbb{E}\left\{ W(t) + M(t) \right\} \\
		\text{Subject to} & \displaystyle
		(\ref{constraint: association}),
		(\ref{constraint: accessibility}),
		(\ref{constraint: queue stability}).
	\end{array}
\end{equation}

\section{Algorithm Design and Performance Analysis}  \label{sec: algorithm design}
Provided that the knowledge of system dynamics is fully known \textit{a priori}, Problem (\ref{problem formulation: 1}) can be solved asymptotically optimally in the long run by applying  Lyapunov optimization techniques\cite{neely2010stochastic}.
However, such knowledge is generally hard to attain in practice. 
For example, the communication and computational costs on switches often remain unknown at the beginning of each time slot.
Instead, they can be revealed or inferred from feedback information after requests being delivered or processed. 
Besides, the statistics of each switch $i$'s accessibility to controllers ${A}_{i}(t)$ are also unknown to the systems.
Faced with such uncertainties, online learning is needed to aid the online control procedure. To this end, we have the following challenges to be addressed. 

The \textit{first} challenge relates to the online learning procedure. 
To minimize system costs under such uncertainties, the designed scheme needs to keep a decent balance between \textit{exploration} and \textit{exploitation}. 
Exploration, \textit{i.e.}, favoring those rarely chosen queue backlogs with uncertain costs, allows switches to learn as much about the costs of different choices as possible. 
However, too much exploration may incur unexpectedly higher costs and lead to performance loss (\textit{a.k.a} regret). 
In contrast, if switches stick to exploitation, \textit{i.e.}, selecting queue backlogs with empirically lowest costs, then they may miss potentially better choices and incur undesired regrets. 

The \textit{second} challenge relates to the online control procedure. 
In particular, the design must carefully handle the non-trivial tradeoff between system cost reduction and queue stability. 
This is because improperly choosing the queue backlogs with empirically lowest costs may overwhelm the corresponding switches/controllers. Moreover, maintaining the tradeoff through a series of online decision making makes the design  even more complicated.

The \textit{third} challenge roots in the interplay between online learning and online control. 
On the one hand, recall that the aim of introducing online learning is to reduce uncertainties of system dynamics and to facilitate the online control procedure.
However, if conducted improperly, online learning may misguide the online control procedure and cause more regrets. 
On the other hand, online control aims to make the best use of available information to conduct effective decision making with instructive feedback information. 
But if poorly designed, online control may incur noisy feedback to online learning and impose adverse effects on learning efficiency. 

Given the above challenges, a carefully integrated design of online control and online learning is needed. 
In fact, we can decompose Problem (\ref{problem formulation: 1}) into a series of matching subproblems over time slots. During each time slot, a subset of connections is chosen between switches and their potential targets (including themselves and accessible controllers), subject to constraints (\ref{constraint: association}) and (\ref{constraint: accessibility}).
Each selected connection has a time-varying weight with a constant mean. The goal is to minimize the time-average total weight over selected matchings across time slots.
Faced with such an online decision-making problem under uncertainty, we switch to investigating Problem (\ref{problem formulation: 1}) from the perspective of Combinatorial Multi-Armed Bandit (CMAB)\cite{li2019combinatorial}.
Below we first introduce the background and settings of multi-armed bandit to motivate our reformulation. Then we reformulate Problem (\ref{problem formulation: 1}) as a CMAB problem with long-term constraints, followed by algorithm design and performance analysis.

\subsection{Motivation for Reformulation}
The multi-armed bandit (MAB) model \cite{lattimore2018bandit} has been extensively adopted to study a wide range of sequential decision-making problems under uncertainty \cite{bouneffouf2019survey}. 
Under the canonical settings of MAB, a player is considered to play a multi-armed bandit over a finite number of rounds. In each round, the player pulls one of the arms and receives a reward that is sampled from some distribution with an unknown mean for the selected arm.  
The objective of MAB model is to devise an online policy for the player to maximize the cumulative reward from its successive plays. Such a model is well suited to scenarios in which each time the decision maker (the player) picks only a single choice. 
Back to our model, a direct idea is that we can view each switch as a distinct player and its accessible controllers as the arms. However, such a reformulation is not feasible under our problem settings due to the following reasons. 

First, the decision making of each switch is not independent of each other. Specifically, each controller is often associated with multiple switches during each time slot (round). Accordingly, the decisions of such switches will jointly change the controller's queue backlog and their subsequent decision making. Second, unlike the settings of MAB model, the set of arms for the decision maker keeps change across different time slots. 
Therefore, the basic MAB model does not apply to our problem settings in the face of the coupled nature of decision making among switches and the time-varying availability of switch-controller connections. 

Instead of considering the problem from the perspective of individual switches, we view the whole data plane as a player and each switch-controller connection as a distinct arm. Then the problem becomes how to pick a set of arms (instead of one arm) for the player during each time slot to maximize its cumulative rewards (\textit{i.e.}, equivalent to the minus of its incurred costs) over time slots. We find that such a setting is well defined by the combinatorial multi-armed bandit (CMAB) model with sleeping arms \cite{li2019combinatorial}. In the following subsections, we formalize the settings of CMAB and demonstrate our problem reformulation in detail.

\subsection{Combinatorial Bandit Setting}
\begin{figure}
  \centering 
  \includegraphics[scale=.36]{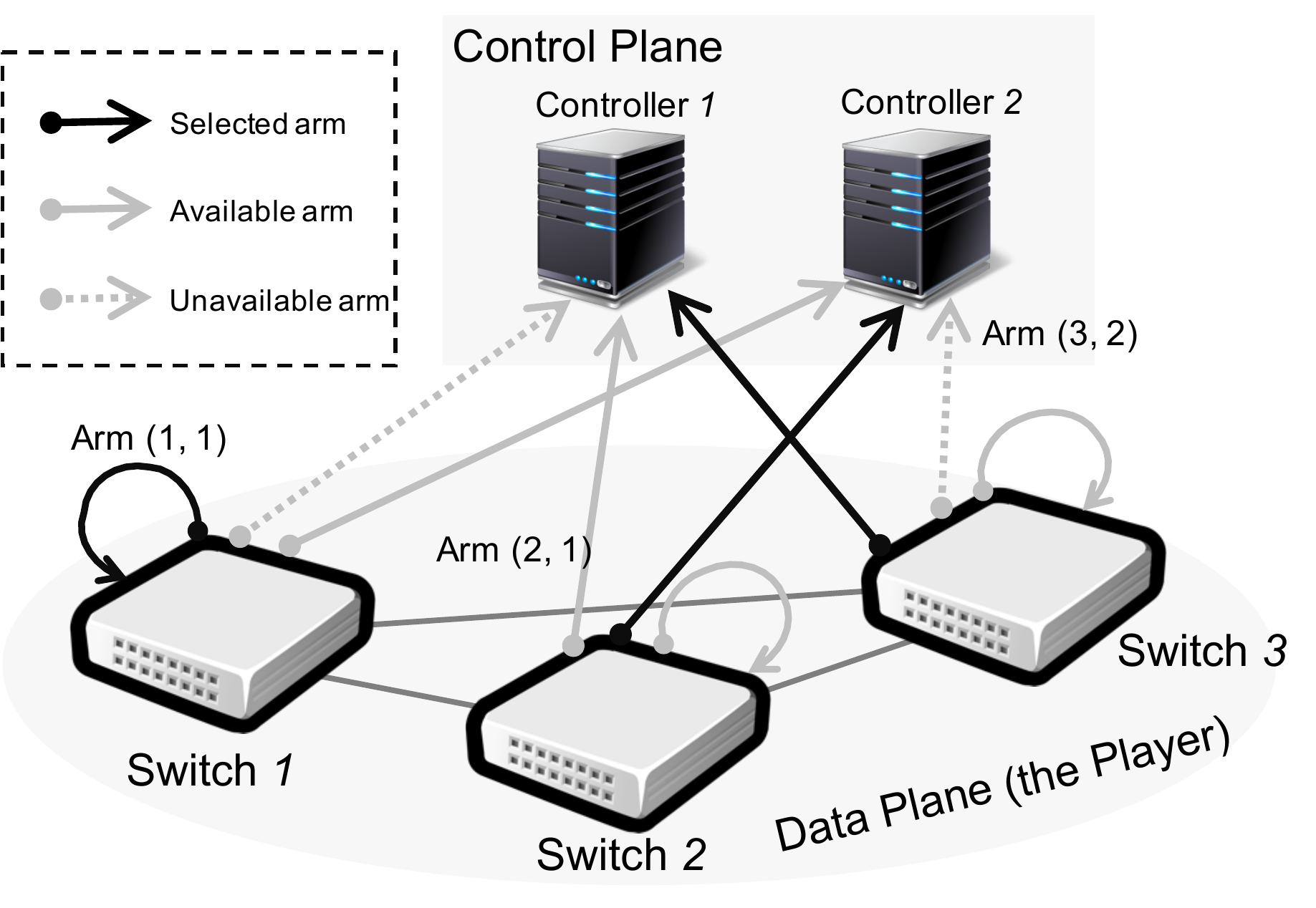}
\caption{An illustration of our problem reformulation. During each time slot, the data plane (as the player) has access to a set of connections between switches and controllers. Each accessible connection is viewed as an available arm (denoted by a solid arrow in the figure), while the other connections are unavailable (each denoted by a dashed arrow). The player aims to pick a subset of arms (\textit{e.g.}, arms $(1,1)$, $(2,2)$, and $(3, 1)$) so as to maximize its cumulative rewards (\textit{i.e.}, minimize the cumulative system costs) in the long run with queue stability guarantee.}
\label{figure: reform-model}
\end{figure}

Suppose that a player plays a bandit with $N$ arms over $T$ time slots.
The set of all arms is denoted by $\mathcal{N} = \{1, 2, \dots, N\}$. 
During each time slot $t$, only a subset of arms are available to the player. 
We denote their arms by set $A(t) \in \mathcal{P}\left( \mathcal{N} \right)$. 
The availability of arms follows a fixed but unknown distribution $p(Z) = \text{Pr}\left\{ A(t) = Z \right\}$ for $Z \in \mathcal{P}( \mathcal{N} )$.
Then given $A(t)$, the player needs to choose and pull a subset of arms $f(t)$ (\textit{a.k.a.} \textit{super arm}) whose size is no more than some constant $m$. 
Accordingly, we define $\mathcal{F}(A(t)) \triangleq \{ f \subseteq A(t): |f| \leq m \}$ as the set of feasible super arms. 
For each arm $n \in f(t)$, it is associated with some reward $X_{n}(t)$ which follows a fixed but unknown distribution with  a mean $\bar{x}_{n}$.
Therefore, after pulling the arms in super arm $f(t)$, the player will receive a compound reward of $R(t) \triangleq \sum_{n \in f(t)} X_{n}(t)$. 
The goal of the player is to find a policy $\pi$ that maximizes the expected time-average compound reward (denoted by $\mathbb{E}\big\{ \frac{1}{T} \sum_{t=0}^{T-1} R(t) \big\}$) over $T$ time slots.
Such a goal is also equivalent to minimizing reward loss (regret) due to decision making under uncertainty. 
The regret is characterized by the difference between the maximum achievable reward given the full knowledge of system dynamics and the time-average expected reward under policy $\pi$, \textit{i.e.},
\begin{equation}\label{regret}
	\begin{array}{c}
		\displaystyle
		R_{\pi}(T) \triangleq R^{*} - \mathbb{E}\Big\{ \frac{1}{T} \sum_{t=0}^{T-1} \sum_{n \in f(t)} X_{n}(t) \Big\},
	\end{array}
\end{equation}
where $R^{*} \triangleq \sum_{Z \in \mathcal{P}(\mathcal{N})} p(Z) \sum_{f \in \mathcal{F}(Z)} q^{*}_{Z}(f) \sum_{n \in f} \bar{x}_{n}$ and $q^{*}_{Z}(f)$ denotes the optimal expected time fraction of choosing super arm $f$ from set $\mathcal{P}(Z)$.
In the following subsection, we conduct problem reformulation by fitting our model into the above settings. An illustration of such a reformulation is given in Figure \ref{figure: reform-model}.

\setlength{\textfloatsep}{3pt}

\subsection{Problem Reformulation}
We regard each potential connection between switches and controllers as an arm and construct the arm set $\mathcal{N}$ as $\mathcal{N} \triangleq \left\{ (i, i) \right\}_{i \in \mathcal{S}} \cup \left\{ (i, j) \right\}_{i \in \mathcal{S}, j \in \mathcal{C}_{i}}$. 
For each arm $(i, i) \in \mathcal{N}$ with $i \in \mathcal{S}$, pulling it corresponds to the decision to make switch $i$ process new requests locally;
for each arm $(i, j) \in \mathcal{N}$ with $i \in \mathcal{S}$ and $j \in \mathcal{C}_{i}$, pulling it corresponds to the decision to associate switch $i$ with controller $j$.
Then we denote the set of all available arms during time slot $t$ by $A(t) \triangleq \left\{ (i, i) \right\}_{i \in \mathcal{S}} \cup \left( \bigcup_{i \in \mathcal{S}} A_{i}(t) \right)$.
Note that such an available arm set is also \textit{i.i.d.} across time slots and independent across switches. 
Then the probability of attaining such an arm set by $p(A(t)) = \prod_{i \in \mathcal{S}} p(A_{i}(t))$. 
Accordingly, we define the set of feasible super arms $\mathcal{F}(A(t))$ by
\begin{equation}\label{definition: feasible set}
	\begin{array}{l}
		\displaystyle
		\mathcal{F}(A(t)) \triangleq \\ 
		\displaystyle
		\{ f \subseteq A(t): 
			\mathds{1}_{(i, i) \in f} + \sum_{j \in \mathcal{C}_{i}} \mathds{1}_{(i, j) \in f} = 1,\  \forall\ i \in \mathcal{S} \},
	\end{array}
\end{equation}
where $\mathds{1}_{n \in f}$ indicates whether arm $n$ is contained in super arm $f$. 
Note that (\ref{definition: feasible set}) 
ensures that each super arm will contain exactly $|\mathcal{S}|$ arms. In other words, given the accessible controller set $A(t)$, 
the constraint in (\ref{definition: feasible set}) is equivalent to constraint (\ref{constraint: association}). 

Next, we define each arm $(i, i)$'s reward during time slot $t$ as $X_{i, i}(t) = -M_{i}(t)$, \textit{i.e.}, the negative of the computational cost of switch $i$.
For each arm $(i, j)$, we use $X_{i, j}(t) = -W_{i, j}(t)$ to denote the reward of associating switch $i$ with controller $j$. 
Given any super arm $f(t)$, its corresponding compound reward $R(t)$ is given by 
\begin{equation}\label{definition: compound reward}
	\begin{array}{c}
		\displaystyle
		R(t) \triangleq \sum_{(i, k) \in f(t)} X_{i, k}(t).
	\end{array}
\end{equation}
Defined in the above way, minimizing the time-average total system costs in (\ref{problem formulation: 1}) is equivalent to maximizing the cumulative rewards over $T$ time slots. 
Thus, Problem (\ref{problem formulation: 1}) can be rewritten as follows
\begin{equation}\label{problem formulation: 2}
	\begin{array}{cl}
		\underset{ \{f(t) \in \mathcal{F}(A(t)) \}_{t=0}^{T-1} }{\text{Maximize}} & \displaystyle
		\frac{1}{T} \sum_{t=0}^{T-1} \mathbb{E}\Big\{ \sum_{(i, k) \in f(t)} X_{i, k}(t) \Big\} \\
		\text{Subject to} & \displaystyle
		(\ref{constraint: queue stability}).
	\end{array}
\end{equation}

\textbf{Remark:}
	Problem (\ref{problem formulation: 2}) is more complicated than the original CMAB problem. 
	Due to constraint (\ref{constraint: queue stability}), the decision-making process should not only maximize the cumulative compound reward but also maintain queue stability on switches and controllers in the long run. 
	This requires the online control procedure not to narrow down its sight onto the arms with the empirically highest rewards; instead, it should also switch amongst multiple arms by taking their queue backlog sizes into account. 
	Note that in \cite{li2019combinatorial}, queue stability is considered as a form of fairness guarantee. Meanwhile, in their model, different arms' queue backlogs are independent of each other. 
	However, in our model, each arm corresponds to a switch-controller connection that directs requests sent from a switch to one of its accessible controllers or itself. 
	Accordingly, each controller's queue is shared by multiple switches (equivalently by multiple arms). Such a coupling among arms makes our problem even more complicated.

\subsection{Algorithm Design}
Regarding online learning, we adopt \textit{UCB1-tuned}\cite{auer2002finite} to estimate each arm's quality based on its feedback reward information. 
As an extended version of classic UCB1 (upper-confidence-bound1) method\cite{agrawal1995sample},
UCB1-tuned makes further use of feedback information to estimate the reward variance of each arm and reconcile the \textit{exploration-exploitation} tradeoff.
Owing to such fine-grained evaluations, UCB1-tuned has been shown to outperform various bandit learning methods\cite{auer2002finite}. 

In particular, for each arm $(i, k)$, we define $h_{i, k}(t)$ as the number of times it has been chosen by the end of time slot $t$. Formally, we write $h_{i, k}(t) \triangleq \sum_{\tau = 1}^{t} I_{i, k}(\tau)$ and assume $h_{i, k}(-1) = 0$. 
Then the sample mean of rewards is given by
\begin{equation}
	\begin{array}{c}
		\displaystyle
		\hat{x}_{i, k}(t) \triangleq \frac{
			\sum_{\tau=1}^{t} X_{i, k}(\tau) I_{i, k}(\tau)
		}{h_{i, k}(t)}.
	\end{array}
\end{equation}
Note that if arm $(i, k)$ has never been chosen before, we set $\hat{x}_{i,k}(t)=0$.

Next, we define the UCB1-tuned estimate for arm $(i, k)$'s mean reward as
\footnote{$\tilde{x}_{i, k}(-1)$ is initialized as zero for all $i \in \mathcal{S}$ and $k \in \{i\} \cup \mathcal{C}_{i}$.}
\begin{equation}\label{equation: update ucb}
	\begin{array}{l}
		\displaystyle
		\tilde{x}_{i, k}(t+1) \triangleq \min\left\{
			u_{i, k}(t), 0
		\right\},
	\end{array}
\end{equation}
where the upper confidence bound $u_{i, k}(t)$ is defined as 
\begin{equation}\label{definition: ucb_bound}
	u_{i, k}(t) \triangleq 
	\hat{x}_{i, k}(t) + 
	\beta
	\sqrt{\frac{\log{(t+1)}}{h_{i, k}(t)} \cdot \min\{ 1/4, V_{i, k}(h_{i, k}(t)) \} },
\end{equation}
and operator $\min\{\cdot, 0\}$ is to ensure the non-positiveness of reward estimates, since system costs are always non-negative.
In (\ref{definition: ucb_bound}), parameter $\beta$ measures the relative weight of uncertainty credit (the second term in (\ref{definition: ucb_bound}), \textit{a.k.a.} the exploration term) 
to the empirical estimate of mean reward $\hat{x}_{i, k}(t)$ (\textit{a.k.a.} the exploitation term). 
Accordingly, different values of parameter $\beta$ reflect different degrees of exploration-exploitation tradeoff. 
Note that we can view the exploration term as the scaled error margin of the reward estimate. Specifically, in the exploration term, $V_{i, k}$ is defined as 
\begin{equation}
	\begin{array}{l}
		\displaystyle
		V_{i, k}(h_{i, k}(t)) \triangleq \\
		\displaystyle
		\Big( \frac{1}{h_{i, k}(t)} \sum_{\tau=1}^{h_{i, k}(t)} \left[X_{i, k}(\tau)\right]^{2} - { \bar{X}_{i, k} }^{2} + 
		\sqrt{ \frac{2\log{(t+1)}}{h_{i, k}(t)} }
		\Big),
	\end{array}
\end{equation}
which denotes an optimistic empirical estimate of the reward variance by pulling arm $(i, k)$.

Regarding online control, we adopt Lyapunov drift techniques\cite{neely2010stochastic} to cope with the tradeoff between system cost minimization and queue stability. 
The key idea is to pull arms greedily based on their instant queue backlog sizes and empirical reward estimates during each time slot. 
Through a series of such online decision making, cumulative reward maximization and long-term queue stability can be properly balanced.

By carefully integrating online learning with online control, we devise an effective Learning-Aided Switch-controller Association and Control Devolution scheme (\textit{LASAC}). We show its pseudo-code in Algorithm \ref{algo} with remarks as follows. 
\begin{algorithm}[!h]
 \caption{Learning-Aided Switch-controller Association and Control Devolution (LASAC)}
 \renewcommand{\algorithmicrequire}{\textbf{Input:}}
 \renewcommand{\algorithmicensure}{\textbf{Output:}}
 \begin{algorithmic}[1] \label{algo}
 	\REQUIRE At the beginning of each time slot $t$, given backlog sizes $\boldsymbol{Q}(t)$ and the set of accessible controllers $A_{i}(t)$ for each switch $i$.
 	\ENSURE A series of decisions $\{ \boldsymbol{I}(t) \}_{t}$ over time horizon $T$.
 	\STATE  
 	\textbf{for} each time slot $t \in \{0, 1, \dots, T-1\}$:
    \STATE \IND  
    \textbf{for} each switch $i \in \mathcal{S}$:
	\STATE \IND \IND  
	\textbf{for} each candidate $k \in \{i\} \cup \mathcal{C}_{i}$:
	\STATE \IND \IND \IND  
	\textbf{if} $h_{i, k}(t - 1) > 0$ \textbf{then}:
	\STATE \IND \IND \IND \IND  
	Update $\tilde{x}_{i, k}(t)$ according to (\ref{equation: update ucb}).
	\STATE \IND \IND \IND  
	\textbf{else}
	\STATE \IND \IND \IND \IND  
	Set $\tilde{x}_{i, k}(t) \leftarrow 0$. 
	\STATE \IND \IND  
	Select candidate $k^{*}$ such that
	$$
		k^{*} \in {\arg\min}_{
			k \in \{i\} \cup A_{i}(t)
		} l_{i, k}(t),
	$$
	\IND \IND
	where $l_{i, k}(t)$ is defined as
	\begin{equation}\label{definition: l}
		l_{i, k}(t) \triangleq \left\{
			\begin{array}{ll}
				\displaystyle
				Q^{S}_{i}(t) - V \cdot \tilde{x}_{i, i}(t) &
				\text{if}\ k = i, \\
				\displaystyle
				Q^{C}_{k}(t) - V \cdot \tilde{x}_{i, k}(t) &
				\text{otherwise.}
			\end{array}
		\right.
	\end{equation}
	\STATE \IND \IND  
	Set $I_{i, k^{*}}(t) \leftarrow 1$ and $I_{i, k}(t) \leftarrow 0$ for $k \neq k^{*}$.
	\STATE \IND \IND  
	\textbf{if} $k^{*} = i$ \textbf{then}
	\STATE \IND \IND \IND  
	Append new requests to switch $i$'s local queue.
	\STATE \IND \IND  
	\textbf{else}
	\STATE \IND \IND \IND  
	Forward new requests to controller $k^{*}$.
	\STATE \IND \IND  
	Collect the corresponding reward $X_{i, k^{*}}(t)$.
	\STATE \IND \IND
	Update $h_{i, k}(t)$ and $\hat{x}_{i, k}(t)$ for all $k$.
	\STATE \IND  
	Update all queue backlogs according to 
	(\ref{queueing dynamics: switches}) and
	(\ref{queueing dynamics: controllers}).
\end{algorithmic}
\end{algorithm}
\setlength{\textfloatsep}{0pt}
\begin{remark}
	By maintaining the estimate (\ref{definition: ucb_bound}) for each arm in every time slot, LASAC well balances the tradeoff between exploration and exploitation for online learning. 
	Particularly, if an arm has been pulled for a sufficiently large number of times, the exploitation term will become dominant and the estimate will count more on the empirical mean. 
	Otherwise, the exploration term comes into play. 
	Recall that $V_{i, k}(h_{i, k}(t))$ is an optimistic estimate of the variance of the arm's reward.
	Then intuitively, under UCB1-tuned, the credit of uncertainty for pulling each arm is large not only when the arm is explored insufficiently but also when its reward variance estimate is high. In this way, exploration is further encouraged in LASAC, which conduces to learning efficiency. 
\end{remark}
\begin{remark}
	According to line $8$ in Algorithm \ref{algo}, LASAC manages the tradeoff between cumulative compound reward maximization and queue stability by jointly considering each arm's reward estimate and its associated instant queue backlog size. 
	By (\ref{definition: l}), LASAC takes instant queue backlog sizes as the indicator of queue stability. 
	Meanwhile, LASAC uses the value of parameter $V$ to determine the relative importance of reward maximization compared to queue stability. 
	The larger the value of $V$ is, the more willing LASAC is to pick the arm with the empirically highest reward (lowest cost) for each switch.
	In contrast, if the value of parameter $V$ is small, then LASAC will favor those arms with small queue backlog sizes. 
	In practice, the value of parameter $V$ can be tuned around the ratio of the magnitude of individual queue backlog capacity to that of system costs.
\end{remark}
\begin{remark}
	LASAC can run in a distributed manner with a computational complexity of $O(|\mathcal{C}|)$. Specifically, given instant queue backlog sizes and accessible states of its accessible controllers, each switch can conduct its own decision making independently. In practice, LASAC can be implemented and deployed in either a centralized or a decentralized manner, which are visualized in Figures 2(a) and 2(b), respectively. 
	
		Under centralized implementation, LASAC can be deployed on a particular server that is independent of both the control plane and the data plane. 
		At runtime, LASAC should collect instant system dynamics including the queue backlog sizes on  switches and controllers, as well as the communication and computation costs, to conduct the decision making. Then it will spread such decisions onto switches, so that switches can schedule their requests accordingly. The advantage of such an implementation is that it requires no modification on switches; \textit{i.e.}, all necessary information for decision making can be obtained via standard OpenFlow APIs\cite{openflowapi}. 
		This is well-suited for scenarios with a large-scale data plane in which switches' compute resources are scarce. 
		However, such an implementation may also be a single point of failure and the performance bottleneck for a large-scale data plane. Besides, it requires extra communication overheads for exchanging information between the system and LASAC.
		
		Under decentralized implementation, LASAC is implemented as a function module and deployed on each switch. Then at runtime, each switch periodically updates its information about system costs and queue backlogs from the control plane, to conduct its decision making independently. Although such a way requires modification on switches, compared to the  centralized implementation, it requires less amounts of information exchange and thus incur lower communication overheads. Moreover, the resulting distributed decision making also conduces to better scalability and fault tolerance.
\end{remark}
\begin{figure}[!t]
  \centering 
  \includegraphics[scale=.265]{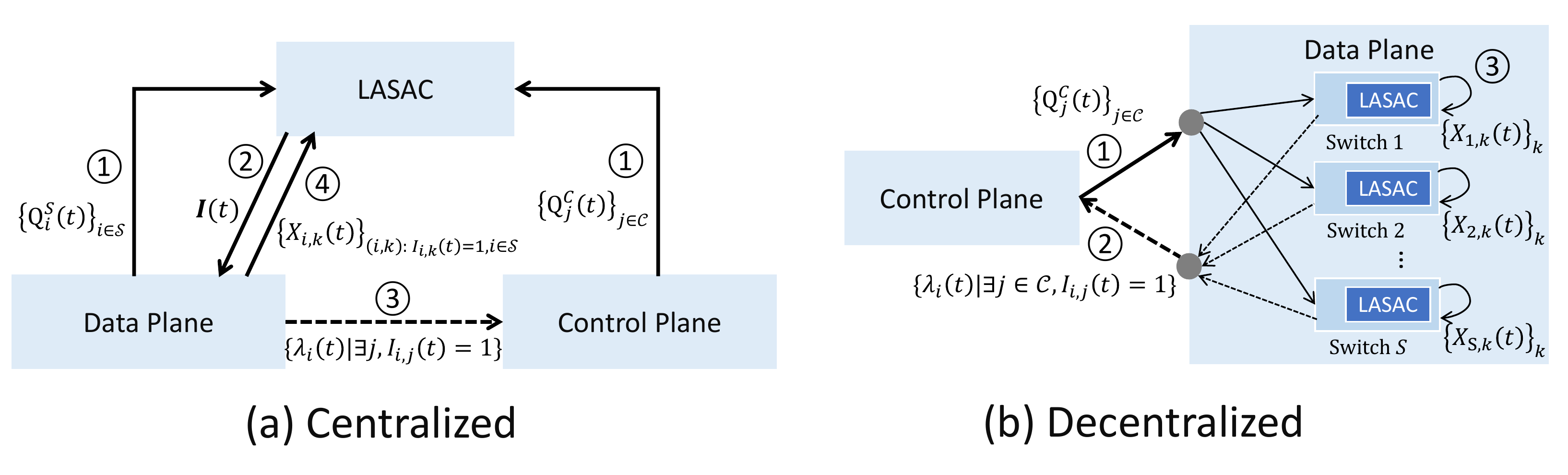}
  \vspace{-0.5cm}
\caption{Two implementations of LASAC}
\label{figure: perf-with-V}
\end{figure}
\setlength{\textfloatsep}{0pt}

\subsection{Performance Analysis}
In general, we have two questions about the performance of LASAC.
One is that, with learning and control procedures being tightly coupled, can LASAC guarantee queue stability in the long run? 
The other is that in the face of various uncertainties, what is the regret upper bound that LASAC can achieve? How would different factors such as the length of time horizon $T$ affect the regret bound? 
These two questions are answered by the following theorems, respectively. 

\textbf{Queue Stability:}
For any mean service rate vector $\boldsymbol{\mu}=( \bar{\mu}^{S}_{1}, \dots, \bar{\mu}^{S}_{|\mathcal{S}|}, \bar{\mu}^{C}_{1}, \dots, \bar{\mu}^{C}_{|\mathcal{C}|} )$, it is said to be \textit{feasible} if there exists such a scheduling scheme that its decision making ensures that the mean arrival rate is no greater than the mean service rate for each switch and controller in the system.  
We define the set of all such feasible mean service rate vectors as the \textit{maximal feasibility region}. We have the following result.
\begin{theorem}\label{theorem1}
	Provided that the mean service rate vector $\boldsymbol{\mu}$ lies in
the interior of the maximal feasibility region, then LASAC can achieve queue stability (\ref{constraint: queue stability}) in the systems. In other words, there exist  positive constants $B$ and $\epsilon$ such that
	\begin{equation}\label{queueing_bound}
	\begin{array}{r}
	\displaystyle
	\limsup _ { T \rightarrow \infty } \frac { 1 } { T } \sum _ { t = 0 } ^ { T - 1 } \mathbb { E } \bigg\{ \sum _ { i \in \mathcal { S } } Q _ { i } ^ { S } ( t ) + \sum _ { j \in \mathcal { C } } Q _ { j } ^ { C } ( t ) \bigg\} \leq \\
	\displaystyle
	\frac { { B } } { \epsilon } 
	 + V \cdot | \mathcal { S } | \cdot \lambda _ { \max } \cdot \max \left\{ w _ { \max } , m _ { \max } \right\}.
	\end{array}
	\end{equation}
\end{theorem}
\textit{Proof Sketch:} 
We leverage Lyapunov drift techniques \cite{neely2010stochastic} to conduct the proof. 
First, we introduce a Lyapunov drift function to characterize the change of queue backlog sizes between successive time slots. 
Then we switch our focus onto the notion of drift-plus-penalty function, which integrates the objective function into the drift function. Based on the assumption that the total service capacity of the system is greater than the total mean traffic rate, we derive an upper-bound on the drift-plus-penalty function, which is independent of any particular policy. Finally, by exploiting telescoping sum techniques, we obtain the desired result. More details of the proof are given in Appendix \ref{proof of theorem 1}.

\textbf{Regret Bound:} The following theorem gives an upper bound for the regret of LASAC over time horizon $T$.
\begin{theorem}\label{theorem2}
Over $T$ time slots, the time-averaged regret (\ref{regret}) of LASAC has an $O(\sqrt{\log{T}/T})$ sublinear upper bound; \textit{i.e.},
	there exists some positive constant $\tilde{B}$ such that
	\begin{equation}\label{regret_bound}
	\begin{aligned}
	&R_{\text{LASAC}}(T) \leq \frac { \tilde{B} } { V } 
	+ { 2 | S | \cdot ( | C | + 1 ) } \cdot \bigg[ \beta \sqrt { \frac{\log{T}}{T}  } + 
	\\
	& \frac{1}{T} ( G_{\beta}
	 + \frac{1}{2} ) \cdot \max\{w_{\text{max}},m_{\text{max}}\} \bigg],
	\end{aligned}
	\vspace{-0.5em}
	\end{equation}
where $G_{\beta} \triangleq \sum_{t = 1}^{\infty} t^{-\frac{\beta^{2}}{2}}$ is a function of parameter $\beta$.
\vspace{0.3em}
\end{theorem}
\textit{Proof Sketch:} 
First, we introduce the notion of per-time-slot regret to measure the difference between the expected rewards that are achieved under the optimal policy and LASAC. Based on such a notion, we derive an upper-bound for the drift-plus-regret term that characterizes the performance difference between the LASAC and an auxiliary policy with a linear function. 
Then by dividing such a bound into three sub-terms, we adopt techniques such as Chernoff-Hoeffding bound and Jensen's inequality to bound each sub-term to complete the proof. 
More details of the proof are given in Appendix \ref{proof of theorem 2}.

\begin{figure*}[!t]
  \centering   
  \hspace*{-1.8cm} 
  \includegraphics[scale=.21]{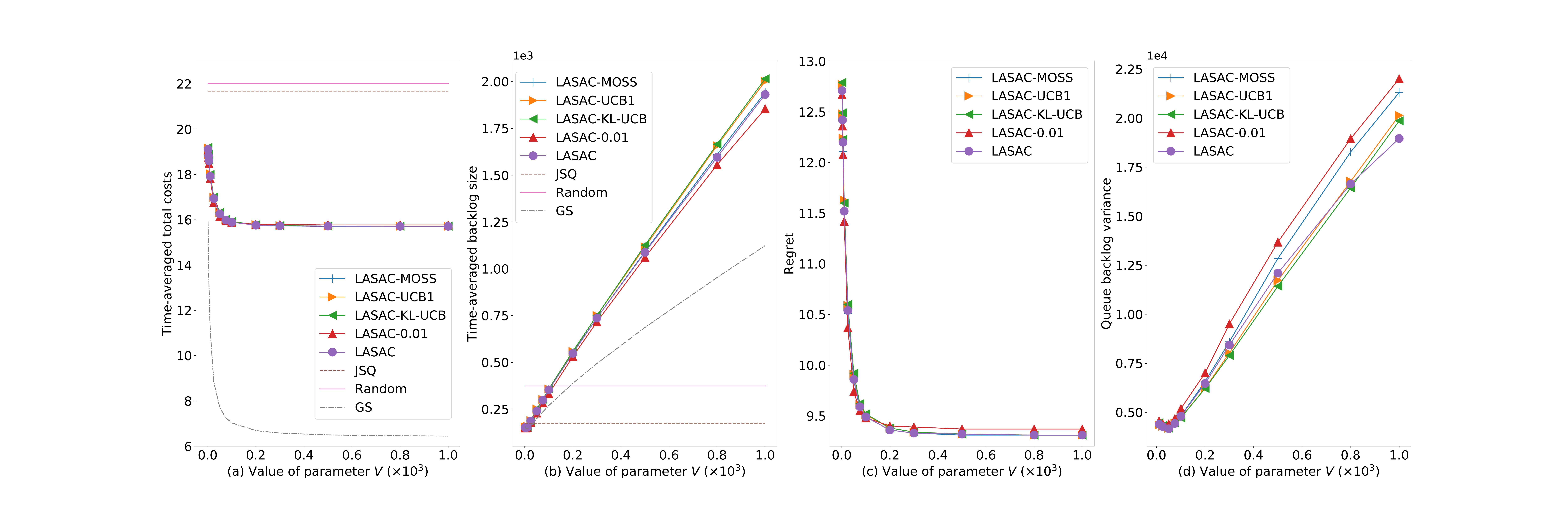}
  \vspace{-0.8cm}
\caption{Performance Comparison under Different Values of $V$ with $\beta=2$}
\label{figure: perf-with-V}
\end{figure*}
\setlength{\textfloatsep}{0pt}

\textbf{Discussion:} From Theorems $1$ and $2$, we know that:
\begin{enumerate}
	\item Regarding online control, LASAC can achieve a tunable $[O(1/V), O(V)]$ tradeoff between system cost reduction (reward maximization) and queue stability in terms of an upper bound for the total queue backlog size. 
	Particularly, according to (\ref{queueing_bound}), the upper bound of queue backlog size is linear in the value of $V$; and according to (\ref{regret_bound}), the regret is inversely proportional to the value of $V$. That being said, larger values of $V$ lead to larger total queue backlog sizes and smaller regrets, whilst smaller values of $V$ result in smaller backlog sizes and larger regrets.
 	\item Regarding online learning, LASAC can achieve different levels of exploration-exploitation tradeoff with different values of $\beta$. Parameter $\beta$ appears in two terms of (\ref{regret_bound}), \textit{i.e.}, $\beta \sqrt { {\log{T}}/{T} }$ and $(G_{\beta} + 1/2) \max\{w_{\text{max}},m_{\text{max}}\} / T$. 
	Recall from (\ref{definition: ucb_bound}) that the greater the value of parameter $\beta$, the more explorations LASAC conducts. 
 	However, the first term in (\ref{regret_bound}) suggests that more explorations incur a higher regret (equivalently high system costs). 
 	Nonetheless, the regret will decrease over time at a rate of $O(\sqrt{\log{T}/T})$.
 	Besides, if the value of parameter $\beta$ is tuned too small, then the series $G_{\beta}$ may not converge and hence the bound in (\ref{regret_bound}) becomes indefinite. This reflects the downside of over-conservative exploration. That is, each switch may blindly choose the controllers with the empirically lowest cost estimates and miss other potentially better candidates. 
 	\item The first and second terms in (\ref{regret_bound}) also quantify the impact of online control and online learning on the regret, respectively. However, there are other subtle interactions between control and learning. For example, the online control procedure would also implicitly enforce exploration, which is discussed in Section \ref{sec: simulation}. 
\end{enumerate}

%

\section{Simulation} \label{sec: simulation}

\subsection{Basic Settings}
We conduct simulations in an SDN system that evolves over $5 \times 10^{6}$ time slots. Each time slot has a length of $10ms$.
The system contains $10$ switches in the data plane and $4$ controllers in the control plane.
Each switch has three potentially connected controllers, with each controller being chosen independently randomly with a probability ranging from $0.8$ to $1$.
Meanwhile, for each switch and controller, we set their mean processing capacities as $2$ and $12$ requests per time slot, respectively.
The mean computation and communication costs of each switch are set equal to the number of cores assigned to process requests and the number of hops to its associated controllers, respectively. 
Besides, request arrivals on each switch follow the distribution drawn from the measurement of real-world networks\cite{benson2010network}. 
As for the value of parameter $V$, recall that by (\ref{definition: l}), it measures the tradeoff between reward maximization (system cost minimization) compared to queue stability. Accordingly, when cost minimization and queue stability are deemed equally important, the value of $V$ is about the ratio of the magnitude of queue backlog size to that of system costs, which is $100$ under our settings. 
In simulations, we adjust the value of $V$ to investigate system performance under different levels of tradeoffs. 
To be specific, we take the value of parameter $V$ from range $[1, 1000]$.
To eliminate the impact of randomness, all results are averaged over $20$ runs.

\textbf{Baseline Schemes:}
We compare LASAC with the following baseline schemes which run on a per-time-slot basis.
\begin{enumerate}
	\item[$\diamond$] Greedy Scheduling (GS)\cite{huang2017dynamic}: Each switch is assumed to have \textit{full knowledge} about all communication and computational costs at the beginning of each time slot. It makes decisions with (\ref{definition: l}) based on the actual information rather than the empirical estimates.
	\item[$\diamond$] Random scheme: Each switch uniformly randomly forwards requests to itself or one of its accessible controllers. 
	\item[$\diamond$] Join-the-shortest-queue (JSQ): Each switch chooses the controller (or itself) with the smallest queue backlog size.  
\end{enumerate}
Besides, we also propose some variants of LASAC, with different ways to handle the exploration-exploitation tradeoff. 
\begin{itemize}
	\item[$\diamond$] LASAC-with-$\epsilon$-Greedy (LASAC-$\epsilon$): With probability $\epsilon$, each switch selects one of its accessible controllers or itself uniformly randomly. Otherwise, the same decision making as LASAC is conducted.
	\item[$\diamond$] LASAC-$X$: Variants of this type follow the same decision making as LASAC, except that switches replace the reward estimate (\ref{definition: ucb_bound}) by other UCB variant $X$, including UCB1\cite{agrawal1995sample}, MOSS\cite{audibert2009minimax}, and KL-UCB\cite{maillard2011finite}.
\end{itemize}

\subsection{Simulation Results and Analysis}
We evaluate the performance of LASAC by comparing it with baseline schemes and investigating how it handles the tradeoff between system cost reduction and long-term queue stability, as well as the interplay between online learning and online control under different parameter settings.

\textbf{Performance under different choices of parameter $V$:} In
Figures \ref{figure: perf-with-V}(a) - \ref{figure: perf-with-V}(d), we set $\beta=2$ and evaluate different schemes as the value of $V$ increases from $1$ to $1000$.\footnote{Note that in (\ref{regret_bound}), the series $G_{\beta}$ converges only when $\beta \in (\sqrt{2}, +\infty)$. When $\beta \leq \sqrt{2}$, the series $G_{\beta}$ diverges and Theorem \ref{theorem2} holds trivially.}

Figures \ref{figure: perf-with-V}(a) and \ref{figure: perf-with-V}(b) show that GS achieves near-optimal total costs as the value of $V$ increases to $1000$. This is owing to its full knowledge about communication and computation costs during each time slot upon decision making. 
In contrast, without such prior information, LASAC and its variants incur higher total costs and larger queue backlog sizes than GS. 
Meanwhile, Random and JSQ, although achieving more balanced workloads with small backlog sizes, achieve higher (up to $40.1\%$) costs than LASAC as they make no use of system cost information. 

More specifically, as the value of $V$ increases from $1$ to $500$, LASAC and its variants lead to $21.6\%$ reduction in system costs and eventually converge thereafter. 
The corresponding regrets of LASAC and its variants are shown in Figure \ref{figure: perf-with-V}(c). 
The cost reduction comes with linear growth in the total queue backlog size and, by \textit{Little's theorem}\cite{little1961proof}, a longer request delay.
As shown in Figure \ref{figure: perf-with-V}(d), 
the increase in the total queue backlog size is mainly due to the decisions made to reduce system costs during the online control procedure. 
For instance, some controllers have lower communication costs to more switches than other controllers. 
Accordingly, they are more likely to become the preferable choices of switches. Recall that by (\ref{definition: l}), the larger the value of $V$ is, the more willing switches are to send new requests to such controllers. 
As a result, such controllers will be loaded with more requests since their service capacities are fixed across all simulations. This demonstrates the increase in the total backlog size. 
Such results verify our theoretical tradeoff between system cost reduction and queue stability in Section V-D.
In practice, the value of $V$ can be tuned to achieve both low costs and small queue backlog sizes (\textit{e.g.}, $V \in [1, 100]$ in our simulations).

\begin{figure}[!t]
  \centering 
  \vspace{-2.4em}
  \includegraphics[scale=.21]{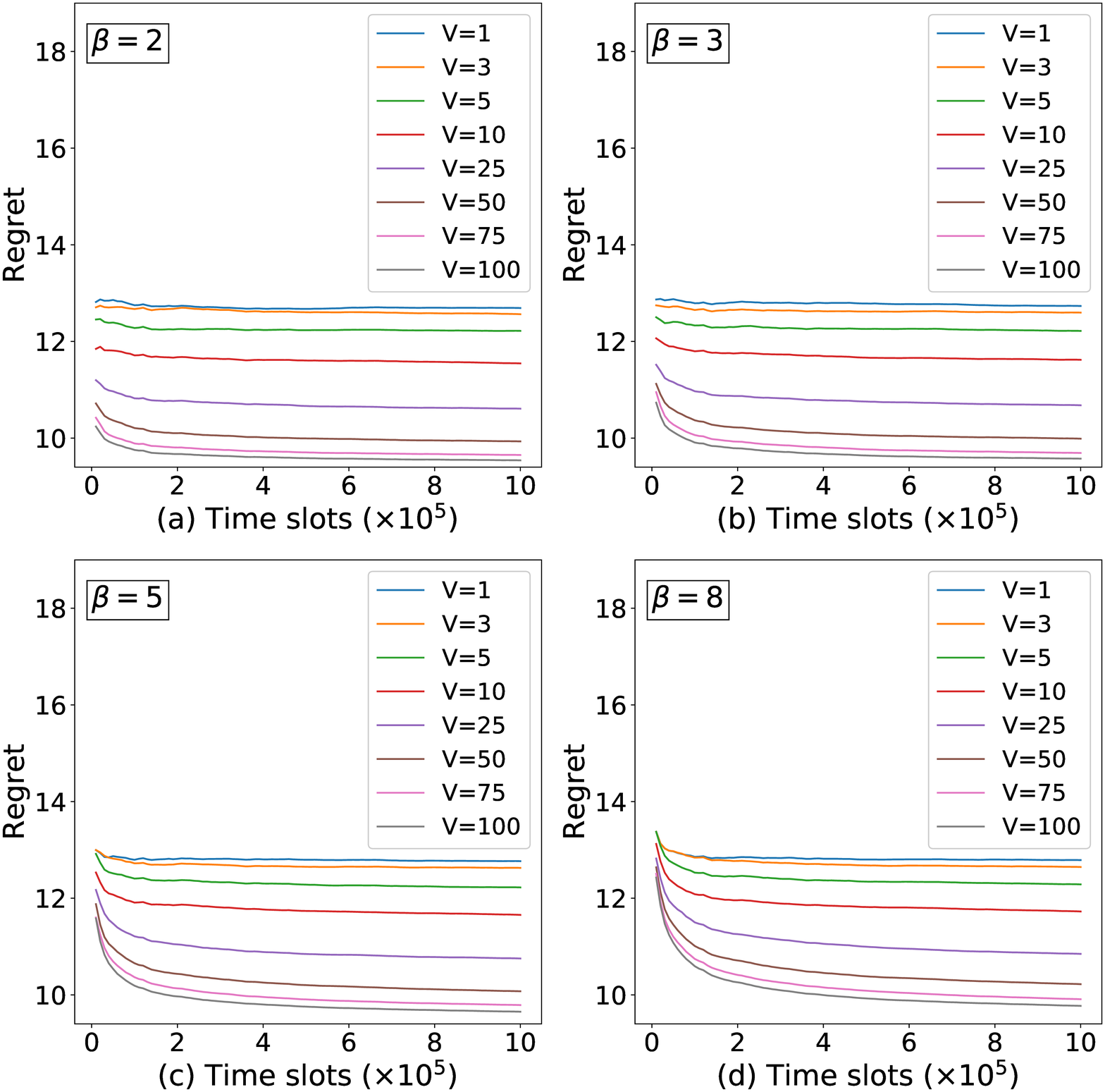}
  \vspace{-1.5em}
   \caption{Regret over Time under Different Settings}
   \label{figure: perf-over-time}
\end{figure}
\setlength{\textfloatsep}{0pt}

\textbf{Regrets over time under different settings}:
In Figure \ref{figure: perf-over-time}, we show how the time-averaged regret of LASAC changes over time under different settings of parameters $V$ and $\beta$. Particularly, we fix the value of $\beta$ as $2$, $3$, $5$, and $8$ in Figures \ref{figure: perf-over-time}(a) - \ref{figure: perf-over-time}(d), respectively. Meanwhile, the values of parameter $V$ vary from $1$ to $100$ in each subfigure. 

From the results, we first see that in general, when the value of $\beta$ is fixed, the regret does not grow linearly over time. Instead, the regret increases initially and then gradually diminishes. The reason is that to achieve effective online learning, each switch needs to conduct sufficient explorations in the early stage. In this process, requests may be forwarded over connections with inferior costs, resulting in a high regret. Then such a regret continues to decrease as each switch learns from more and more feedback information to improve its decision making under LASAC. 
Nonetheless, the regret will not approach zero since the decision-making process should also ensure the stability of queue backlogs in the system. 
Besides, we also see that given the value of $\beta$, the greater the value of parameter $V$, the lower the regret. This implies that increasing the value of $V$ conduces to regret reduction.
Such results verify our results in Theorem \ref{theorem2}.

When it comes to the regrets under different settings of parameter $\beta$, we see that the larger the value of $\beta$, the greater the regret in the early stage. For the reason, recall that the value of $\beta$ measures the score of exploration in (\ref{definition: ucb_bound}). As a result, under LASAC, each switch performs more proactively explorations and incurs a higher regret. Later, such regret curves descend rapidly since more explorations conduce to learning efficiency. 
The lesson learned is that the value of $\beta$ should not be set too large; otherwise, the resulting over-explorations would adversely effect system performance, offsetting the benefits of exploration.

\begin{figure}[!t]
  \centering 
  \includegraphics[scale=.28]{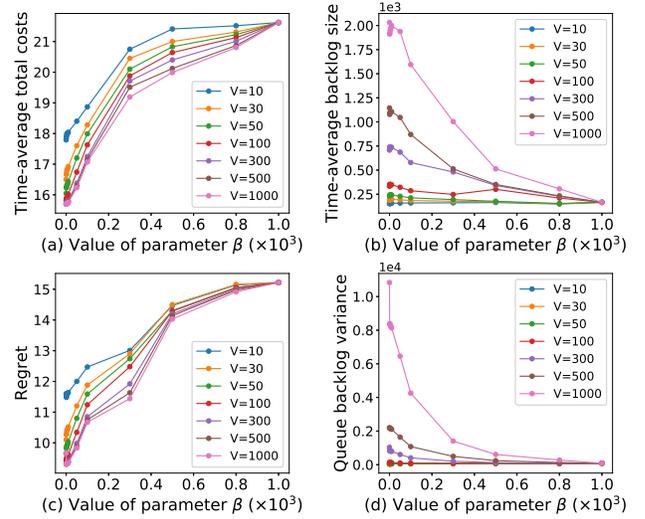}
  \vspace{-0.1cm}
\caption{Time-averaged Performance under Different Values of $\beta$}
\label{figure: perf-with-beta}
\end{figure}

\textbf{The interplay between parameters $V$ and $\beta$:}
To investigate the interplay between online control and online learning, we evaluate the performance of LASAC under various combinations of parameters $V$ and $\beta$. We vary their values in ranges $[10, 1000]$ and $[2, 1000]$, respectively.
Besides, we also evaluate the performance of LASAC given $\beta=0$, which by (\ref{definition: ucb_bound}) corresponds to the strategy with pure exploitation. All results are shown in Figures \ref{figure: perf-with-beta}(a) - \ref{figure: perf-with-beta}(d).

First, we focus on the impact of parameter $\beta$ on system performance. 
Figures \ref{figure: perf-with-beta}(a) and \ref{figure: perf-with-beta}(b) show that larger values of $\beta$ generally lead to higher total system costs (correspondingly larger regrets, as shown in Figure \ref{figure: perf-with-beta}(c)) and smaller total queue backlog sizes. 
Intuitively, according to (\ref{definition: ucb_bound}), 
parameter $\beta$ measures the balance between exploration and exploitation. 
The greater the value of $\beta$ is, the more exploration LASAC intends to conduct. 
Therefore, a large value of $\beta$ results in more frequent exploration and hence excessive system costs. 
In the meantime, frequent exploration prevents LASAC from sticking to forwarding requests to some particular controllers or processing them locally. As shown in Figure \ref{figure: perf-with-beta}(d), this will lead to more balanced queue backlogs across controllers and switches, or equivalently, smaller queue backlog variance. 

\setlength{\textfloatsep}{0pt}

Interestingly, we find that when $\beta=0$, LASAC performs as comparably well as in the cases with $\beta \in [2, 10]$. 
Intuitively, under such a policy with pure exploitation, switches would stick to choosing controllers with initially high empirical estimates and ignore other potentially better candidates, thereby incurring high system costs. 
However, owing to the interplay between online control and online learning, LASAC still maintains a balance between exploration and exploitation even when $\beta=0$.
In fact, LASAC may conduct pure exploitation initially. 
But as soon as those frequently chosen controllers become overloaded due to pure exploitation, LASAC would have the switches turn to other controllers. In other words, exploration is implicitly enforced by the online control procedure to ensure queue stability.

From Figure \ref{figure: perf-with-V} we know that larger values of $V$ lead to lower system costs and larger total backlog size. 
Nevertheless, Figure \ref{figure: perf-with-beta} shows that as the value of parameter $\beta$ increases, the impact of parameter $V$ becomes gradually insignificant. Eventually when $\beta=1000$, \textit{i.e.}, LASAC is forced to make massive explorative decisions such that each switch frequently forwards new requests back and forth among controllers.
As a result, the role of online control becomes marginalized. 
This demonstrates another interplay between online control and online learning. 
In practice, the choice of parameter $\beta$ depends on the tradeoffs in system design. 

\section{Conclusion}  \label{sec: conclusion}
In this paper, we proposed the first scheme with an integrated design of online learning and online control to jointly conduct switch-controller association and control devolution in SDN systems.
Our theoretical analysis showed that LASAC can achieve a tunable tradeoff between queue stability and system cost reduction while ensuring a sublinear time-averaged regret bound over a finite time horizon.
We conducted extensive simulations to verify such theoretical results and evaluate the performance of LASAC and its variants.




\appendices

\section{Proof of Theorem \ref{theorem1}} \label{proof of theorem 1}
We show that LASAC can achieve queueing stability (\ref{constraint: queue stability}) by applying Lyapunov drift analysis techniques\cite{neely2010stochastic}.

First, we define 
$\boldsymbol { Q } ( t ) \triangleq \left\{ Q _ { i } ^ { S } ( t ) \right\} _ { i \in \mathcal { S } } \cup \left\{ Q _ { j } ^ { C } ( t ) \right\} _ { j \in \mathcal { C } }$
as the set of all queue backlogs in the system during time slot $t$. 
Then we consider the following Lyapunov function:
\begin{equation}\label{definition: Lyapunov-function-Q}
L ( \boldsymbol { Q } ( t ) ) \triangleq \frac { 1 } { 2 } \sum _ { i \in \mathcal { S } } \left( Q _ { i } ^ { S } ( t ) \right) ^ { 2 } + \frac { 1 } { 2 } \sum _ { j \in \mathcal  { C } } \left( Q _ { j } ^ { C } ( t ) \right) ^ { 2 }.
\end{equation}
Then we have
\begin{equation}
\begin{aligned}
\nonumber
&\Delta _ { t }  \triangleq L ( \boldsymbol { Q } ( t + 1 ) ) - L ( \boldsymbol { Q } ( t ) )\\
& \stackrel{(a)} = \!\frac { 1 } { 2 } \sum _ { i \in \mathcal { S } } \!\left\{ \! \left( \!\left[ Q _ { i } ^ { S } ( t ) \!+\!\!I _ { i , i } ( t ) \lambda _ { i } ( t ) - \mu _ { i } ^ { S } ( t ) \right] ^ { + } \right) ^ { 2 } \!\!\!- \left( Q _ { i } ^ { S } ( t ) \right) ^ { 2 } \right\} \\
\end{aligned}
\end{equation}
\begin{equation}
\begin{aligned}
& + \frac { 1 } { 2 }\! \sum _ { j \in \mathcal { C } } \bigg\{ \bigg( \bigg[ Q _ { j } ^ { C } ( t )\! + \!\!\sum _ { i \in \mathcal { S } }\! I _ { i , j } ( t ) \lambda _ { i } ( t )\! - \!\mu _ { j } ^ { C } ( t ) \!\bigg] ^ { + }\! \bigg) ^ { 2 } \! \!\!\! - \!\bigg( Q _ { j } ^ { C } ( t ) \bigg) ^ { 2 } \!\bigg\}\\
&\stackrel{(b)} \leq \frac { 1 } { 2 } \sum _ { i \in \mathcal { S } } \bigg\{ \bigg( Q _ { i } ^ { S } ( t ) + I _ { i , i } ( t ) \lambda _ { i } ( t ) - \mu _ { i } ^ { S } ( t ) \bigg) ^ { 2 } - \bigg( Q _ { i } ^ { S } ( t ) \bigg) ^ { 2 } \bigg\} \\
& + \frac { 1 } { 2 } \!\sum _ { j \in \mathcal { C } } \bigg\{ \bigg( Q _ { j } ^ { C } ( t ) \!+ \!\!\sum _ { i \in \mathcal { S } }\! I _ { i , j } ( t ) \lambda _ { i } ( t )\! -\! \mu _ { j } ^ { C } ( t ) \!\!\bigg) ^ { 2 } \!\!\!- \bigg( Q _ { j } ^ { C } ( t ) \bigg) ^ { 2 } \bigg\}\\
& \stackrel{(c)}  \leq \frac { 1 } { 2 } \sum _ { i \in \mathcal { S } } \bigg[ \bigg( I _ { i , i } ( t ) \lambda _ { i } ( t ) \bigg) ^ { 2 } + \bigg( \mu _ { i } ^ { S } ( t ) \bigg) ^ { 2 } \bigg]\\
& + \frac { 1 } { 2 } \sum _ { j \in \mathcal { C } } \bigg[ \bigg( \sum _ { i \in \mathcal { S } } I _ { i , j } ( t ) \lambda _ { i } ( t ) \bigg) ^ { 2 } + \bigg( \mu _ { j } ^ { C } ( t ) \bigg) ^ { 2 } \bigg]\\
& + \sum _ { i \in \mathcal { S } } Q _ { i } ^ { S } ( t ) \bigg( I _ { i , i } ( t ) \lambda _ { i } ( t ) - \mu _ { i } ^ { S } ( t ) \bigg)\\
& + \sum _ { j \in \mathcal { C } } Q _ { j } ^ { C } ( t ) \bigg( \sum _ { i \in \mathcal { S } } I _ { i , j } ( t ) \lambda _ { i } ( t ) - \mu _ { j } ^ { C } ( t ) \bigg)\\
& =  \underbrace{\frac { 1 } { 2 } \sum _ { i \in \mathcal { S } } \bigg( \mu _ { i } ^ { S } ( t ) \bigg) ^ { 2 } + \frac { 1 } { 2 } \sum _ { j \in \mathcal { C } } \bigg( \mu _ { j } ^ { C } ( t ) \bigg) ^ { 2 }}_{U_1(t)} \\
& + \underbrace{\frac { 1 } { 2 } \sum _ { i \in \mathcal { S } } \bigg( I _ { i , i } ( t ) \lambda _ { i } ( t ) \bigg) ^ { 2 } + \frac { 1 } { 2 } \sum _ { j \in \mathcal { C } } \bigg( \sum _ { i \in \mathcal { S } } I _ { i , j } ( t ) \lambda _ { i } ( t ) \bigg) ^ { 2 }}_{U_2(t)}\\
& + \sum _ { i \in S } Q _ { i } ^ { S } ( t ) \bigg( I _ { i , i } ( t ) \lambda _ { i } ( t ) - \mu _ { i } ^ { S } ( t ) \bigg) \\
& + \sum _ { j \in \mathcal { C } } Q _ { j } ^ { C } ( t ) \bigg( \sum _ { i \in \mathcal { S } } I _ { i , j } ( t ) \lambda _ { i } ( t ) - \mu _ { j } ^ { C } ( t ) \bigg), \\
\end{aligned}
\end{equation}
where equality $(a)$ holds since (\ref{queueing dynamics: switches}) and (\ref{queueing dynamics: controllers}), inequality $(b)$ is from the fact that for any real number $x , \big( [ x ] ^ { + } \big) ^ { 2 } = ( \max \{ 0 , x \} ) ^ { 2 } \leq x ^ { 2 }$ and inequality $(c)$ holds because for any two real numbers $x , y \geq 0 ,$ we have $( x - y ) ^ { 2 } \leq x^2 + y^2$.

We define $U_1(t)$ and $U_2(t)$.
\begin{equation}
\begin{aligned}
U_1(t) & \triangleq \frac { 1 } { 2 } \sum _ { i \in \mathcal { S } } \bigg( \mu _ { i } ^ { S } ( t ) \bigg) ^ { 2 } + \frac { 1 } { 2 } \sum _ { j \in \mathcal { C } } \bigg( \mu _ { j } ^ { C } ( t ) \bigg) ^ { 2 }\\
& \stackrel{(a)}\leq \frac { 1 } { 2 } ( | \mathcal { S } | + | \mathcal { C } | ) \cdot \mu _ { \text{max} } ^ { 2 },
\end{aligned}
\end{equation}
where $(a)$ is due to that all service capacities $\mu _ { i } ^ { S } ( t )$ and $\mu _ { j } ^ { C } ( t )$ is upper bounded by constant $\mu _ { \text{max} }$, \textit{i.e.},
\begin{equation}
\begin{aligned}
U_2(t) & \triangleq \frac { 1 } { 2 } \sum _ { i \in \mathcal { S } } \bigg( I _ { i , i } ( t ) \lambda _ { i } ( t ) \bigg) ^ { 2 } + \frac { 1 } { 2 } \sum _ { j \in \mathcal { C } } \bigg( \sum _ { i \in \mathcal { S } } I _ { i , j } ( t ) \lambda _ { i } ( t ) \bigg) ^ { 2 } \\
& \stackrel{(a)} = \frac { 1 } { 2 } \sum _ { i \in \mathcal { S } } \bigg( \lambda _ { i } ( t ) \bigg) ^ { 2 }
\stackrel{(b)} \leq \frac { 1 } { 2 } | \mathcal { S } | \bigg( \lambda _ { \text{max} } \bigg) ^ { 2 },
\end{aligned}
\end{equation}
where $(a)$ holds because of (\ref{constraint: association}) and $(b)$ is by the fact $\lambda _ { i }$ is upper bounded by constant $\lambda _ { \text{max} }$.

Next, we define 
\begin{equation}\label{definition: B}
	B \triangleq \frac { 1 } { 2 } ( | \mathcal { S } | + | \mathcal { C } | ) \cdot \mu _ { \text{max} } ^ { 2 } + \frac { 1 } { 2 } | \mathcal { S } | \bigg( \lambda _ { \text{max} } \bigg) ^ { 2 },
\end{equation}
and we have
\begin{equation}\label{definition:delta-t}
\begin{aligned}
\Delta _ { t } &\leq B + \sum _ { i \in \mathcal { S } } Q _ { i } ^ { S } ( t ) \bigg( I _ { i , i } ( t ) \lambda _ { i } ( t ) - \mu _ { i } ^ { S } ( t ) \bigg) \\
& + \sum _ { j \in \mathcal { C } } Q _ { j } ^ { C } ( t ) \bigg( \sum _ { i \in \mathcal { S } } I _ { i , j } ( t ) \lambda _ { i } ( t ) - \mu _ { j } ^ { C } ( t ) \bigg)\\
& = B - \sum _ { i \in \mathcal { S } } Q _ { i } ^ { S } ( t ) \mu _ { i } ^ { S } ( t ) - \sum _ { j \in \mathcal { C } } Q _ { j } ^ { C } ( t ) \mu _ { j } ^ { C } ( t )\\
& + \sum _ { i \in \mathcal { S } } Q _ { i } ^ { S } ( t ) I _ { i , i } ( t ) \lambda _ { i } ( t ) + \sum _ { j \in \mathcal { C } } \bigg\{ Q _ { j } ^ { C } ( t ) \sum _ { i \in \mathcal { S } } I _ { i , j } ( t ) \lambda _ { i } ( t ) \!\bigg\} \\
& = B - \sum _ { i \in \mathcal { S } } Q _ { i } ^ { S } ( t ) \mu _ { i } ^ { S } ( t ) - \sum _ { j \in \mathcal { C } } Q _ { j } ^ { C } ( t ) \mu _ { j } ^ { C } ( t )\\
& + \sum _ { i \in \mathcal { S } } \lambda _ { i } ( t ) \bigg[ Q _ { i } ^ { S } ( t ) I _ { i , i } ( t ) + \sum _ { j \in \mathcal { C } } Q _ { j } ^ { C } ( t ) I _ { i , j } ( t ) \bigg]\\
& \stackrel{(a)} \leq B - \sum _ { i \in \mathcal { S } } Q _ { i } ^ { S } ( t ) \mu _ { i } ^ { S } ( t ) - \sum _ { j \in \mathcal { C } } Q _ { j } ^ { C } ( t ) \mu _ { j } ^ { C } ( t ) \\
& + \sum _ { i \in \mathcal { S } } \lambda _ { i } ( t ) \bigg[ I _ { i , i } ( t ) \bigg( Q _ { i } ^ { S } ( t ) - V \cdot \tilde { x } _ { i , i } ( t ) \bigg) \bigg] \\
& + \sum _ { i \in \mathcal { S } } \lambda _ { i } ( t ) \bigg[\sum _ { j \in \mathcal { C } } I _ { i , j } ( t ) \bigg( Q _ { j } ^ { C } ( t ) - V \cdot \tilde { x } _ { i , j } ( t ) \bigg) \bigg],
\end{aligned}
\end{equation}
where $(a)$ holds due to (\ref{definition: l}) and $\tilde { x } _ { i , k } ( t )$ is non-positive.

Next we denote the mean vector across all service capacities by $\hat { \boldsymbol { \mu } } \triangleq \Big( \mathbb { E } \Big[ \mu _ { i } ^ { S } \Big] , \ldots , \mathbb { E } \Big[ \mu _ { j } ^ { C } \Big] \Big)$. 
Such a rate vector is said to be \textit{feasible} if there exists a policy that makes decisions over time slots to ensure the stability constraint (\ref{constraint: queue stability}). We denote the set of all such feasible vectors by feasible region $\Gamma$. 
Following this definition, we assume that the vector $\hat { \mu }$ is strictly within region $\Gamma$, 
thus there exists a stationary and randomized policy $\alpha$, characterized by a probability distribution $\boldsymbol { q } = \bigg[ q _ { f } ( Z ) , \forall f \in \mathcal { F } ( Z ) , \forall Z \in \mathcal { P } ( \mathcal { N } ) \bigg]$, 
such that $\sum _ { f \in \mathcal { F } ( Z ) } q _ { f } ( Z ) = 1$ while satisfying the queue stability guarantee (\ref{constraint: queue stability}). 
Particularly, under such a policy, given set of available super arms, $\mathcal { F } ( Z )$, policy $\alpha$ picks a super arm $f$ with probability $q _ { f } ( Z )$ independently across time slots; then for policy $\alpha$, there exists some positive constant $\epsilon$ such that
\begin{equation}\label{instance}
\!\!\!\!\!\sum _ { Z \in \mathcal { P } ( Z ) } \!\!\!\!p ( Z )\! \bigg\{\! \sum _ { f \in \mathcal { F } ( Z ) }\!\!\! q _ { f } ^ { \alpha } ( Z )\!\!\!\!\! \sum _ { i : ( i , k ) \in f } \!\!\!\!\overline { \lambda } _ { i } \!\bigg\} \leq \hat { \mu } _ { k }\! - \!\epsilon , \!\forall k \in \mathcal { S } \cup \mathcal { C }.
\end{equation}
Taking condition expectation of (\ref{definition:delta-t}) given $\boldsymbol { Q } ( t )$, we have
\begin{equation}\label{E-3}
\begin{array}{l}
\displaystyle
\mathbb { E } [ L ( \boldsymbol { Q } ( t + 1 ) ) - L ( \boldsymbol { Q } ( t ) ) | \boldsymbol { Q } ( t ) ] \\
\displaystyle
\leq B - \sum _ { i \in \mathcal { S } } Q _ { i } ^ { S } ( t ) \cdot \overline { \mu } _ { i } - \sum _ { j \in \mathcal { C } } Q _ { j } ^ { C } ( t ) \cdot \overline { \mu } _ { j }+ \\
\mathbb { E } \bigg\{ \sum _ { i \in \mathcal { S } } \lambda _ { i } ( t ) \bigg[ I _ { i , i } ( t ) \bigg( Q _ { i } ^ { S } ( t ) - V \cdot \tilde { x } _ { i , i } ( t ) \bigg)+\\
\displaystyle
\sum _ { j \in \mathcal { C } } I _ { i , j } ( t ) \bigg( Q _ { j } ^ { C } ( t ) - V \cdot \tilde { x } _ { i , j } ( t ) \bigg) \bigg] \bigg| \boldsymbol { Q } ( t ) \bigg\}.
\end{array}
\end{equation}
Then we calculate the
\begin{equation}\label{E-2}
\begin{array}{l}
\nonumber
\displaystyle
\mathbb { E } \bigg\{ \sum _ { i \in \mathcal { S } } \lambda _ { i } ( t ) \bigg[ I _ { i , i } ( t ) \bigg( Q _ { i } ^ { S } ( t ) - V \cdot \tilde { x } _ { i , i } ( t ) \bigg)\\
\displaystyle
+ \sum _ { j \in \mathcal { C } } I _ { i , j } ( t ) \bigg( Q _ { j } ^ { C } ( t ) - V \cdot \tilde { x } _ { i , j } ( t ) \bigg) \bigg] \bigg| \boldsymbol { Q } ( t ) \bigg\}\\
\displaystyle
= \mathbb { E } \bigg\{ \mathbb { E } \bigg\{ \sum _ { i \in \mathcal { S } } \lambda _ { i } ( t ) \bigg[ I _ { i , i } ( t ) \bigg( Q _ { i } ^ { S } ( t ) - V \cdot \tilde { x } _ { i , i } ( t ) \bigg) + \\
\displaystyle
\sum _ { j \in \mathcal { C } } I _ { i , j } ( t ) \bigg( Q _ { j } ^ { C } ( t ) - V \cdot \tilde { x } _ { i , j } ( t ) \bigg) \bigg] \bigg| \boldsymbol { Q } ( t ) , A ( t ) \bigg\} \bigg| \boldsymbol { Q } ( t ) \bigg\}\\
\displaystyle
\stackrel{(a)} \leq \mathbb { E } \bigg\{ \mathbb { E } \bigg\{ \sum _ { i \in \mathcal { S } } \lambda _ { i } ( t ) \bigg[ q _ { i , i } ^ { \alpha } ( A ( t ) ) \bigg( Q _ { i } ^ { S } ( t ) - V \cdot \tilde { x } _ { i , i } ( t ) \bigg)\\
\displaystyle
+ \sum _ { j \in \mathcal { C } } q _ { i , j } ^ { \alpha } ( A ( t ) ) \bigg( Q _ { j } ^ { C } ( t ) - V \cdot \tilde { x } _ { i , j } ( t ) \bigg) \bigg] \bigg| \boldsymbol { Q } ( t ) , A ( t ) \Bigg \} \bigg| \boldsymbol { Q } ( t ) \Bigg\}\\
\displaystyle
= \mathbb { E } \bigg\{ \mathbb { E } \bigg\{ \sum _ { i \in \mathcal { S } }\lambda _ { i } ( t ) \bigg(\sum _ { k \in \{ i \} \cup \mathcal { C } } q _ { i , k } ^ { \alpha } ( A ( t ) ) \bigg[ Q _ { k } ( t ) \\
\displaystyle
-V \!\!\cdot \!\tilde { x } _ { i , k }\! ( t ) \bigg]\Bigg) \bigg| \boldsymbol { Q } ( t ) , A ( t ) \Bigg\} \bigg| \boldsymbol { Q } ( t ) \Bigg\}\\
\displaystyle
\stackrel{(b)} \leq \mathbb { E } \bigg\{ \mathbb { E } \bigg\{ \sum _ { i \in \mathcal { S } } \lambda _ { i } ( t ) \bigg( \sum _ { k \in \{ i \} \cup \mathcal { C } }q _ { i , k } ^ { \alpha } ( A ( t ) ) \bigg[ Q _ { k } ( t ) \\
\displaystyle
- V \cdot { x } _ { \text{min} } \bigg]\bigg) \bigg| \boldsymbol { Q } ( t ) , A ( t ) \bigg\}\bigg| \boldsymbol { Q } ( t ) \Bigg\}\\
\displaystyle
\stackrel{(c)} = \leq \mathbb { E } \bigg\{ \mathbb { E } \bigg\{ \sum _ { i \in \mathcal { S } } \lambda _ { i } ( t ) \bigg( \sum _ { k \in \{ i \} \cup \mathcal { C } }q _ { i , k } ^ { \alpha } ( A ( t ) ) \bigg[ Q _ { k } ( t ) \\
\end{array}
\end{equation}
\begin{equation}\label{E-2}
\begin{array}{l}
\nonumber
\displaystyle
- V \cdot { x } _ { \text{min} } \bigg]\bigg) \bigg| \boldsymbol { Q } ( t ) \bigg\}\bigg| \boldsymbol { Q } ( t ) \Bigg\}\\
\displaystyle
\stackrel{(d)} \leq \mathbb { E } \bigg\{ \mathbb { E } \bigg\{ \sum _ { i \in \mathcal { S } } \lambda _ { i } ( t ) \bigg( \sum _ { k \in \{ i \} \cup \mathcal { C } } 
	\!\!\!
	q _ { i , k } ^ { \alpha }( A ( t ) )  Q _ { k } ( t )\bigg) \bigg|\!A ( t ) \bigg\} \bigg| \boldsymbol { Q } ( t ) \bigg\} \\
\displaystyle
- \!\!\sum _ { i \in \mathcal { S } } \!\!\lambda _ { \text{max} }  V  x _ { \text{min} }\\
\end{array}
\end{equation}
\begin{equation}
= \mathbb { E } \bigg\{ \!\!\sum _ { i \in \mathcal { S } }\!\! \lambda _ { i } ( t ) \bigg( \!\sum _ { k \in \{ i \} \cup \mathcal { C } } \!\!\!\!q _ { i , k } ^ { \alpha }\! ( \!A ( t )\! ) Q _ { k } ( t )\bigg) \bigg| \boldsymbol { Q } ( t ) \bigg\} \!\!- \!\! V |S| \lambda _ { \text{max} }x _ { \text{min} },\\
\end{equation}
where we define 
\begin{equation}\label{definition: x_min}
x_{\text{min}} \triangleq -\max\{w_{\text{max}},m_{\text{max}}\}.
\end{equation} Therein $(a)$ is due to the arg-min operator taken by the algorithm (\ref{definition: l}) during each time slot $t$. On the right-hand-side of inequality $(a)$, $q _ { i , k } ^ { \alpha } ( A ( t ) )$ refers to the probability that arm $(i, k)$ is chosen under policy $\alpha$, $i.e. q _ { i , k } ^ { \alpha } ( A ( t ) ) = \sum _ { f \in \mathcal { F } ( A ( t ) ) : ( i , k ) \in f } q _ { f } ^ { \alpha } ( A ( t ) )$. $(b)$ holds because $\tilde { x } _ { i , k } ( t )$ has lower bound $x_{\text{min}}$. Then $(c)$ is yielded based on the fact that policy $\alpha$'s decision making is independent of the queue backlogs in each time slot. Inequality $(d)$ holds due to the boundedness of $\lambda_{i}(t)$.

We define indicator $\mathds { 1 } _ { ( i , j ) \in f }$ to refer to whether $( i , j ) \in f$ or not. We have
\begin{equation}\label{E-1}
\begin{aligned}
\nonumber
&\mathbb { E } \bigg\{ \sum _ { i \in \mathcal { S } } \lambda _ { i } ( t ) \bigg( \sum _ { k \in \{ i \} \cup \mathcal { C } } q _ { i , k } ^ { \alpha } ( A ( t ) ) Q _ { k } ( t ) \bigg) | \boldsymbol { Q } ( t ) \bigg\}\\
& = \mathbb { E } \bigg\{ \sum _ { i \in \mathcal { S } } \sum _ { k \in \{ i \} \cup \mathcal { C } } \sum _ { f \in \mathcal { F } ( A ( t ) ) : ( i , k ) \in f } \!\!\!\!\!\!\!\!\!\!\!\!\!\!\!\!\!\lambda _ { i } ( t ) q _ { f } ^ { \alpha } ( A ( t ) ) Q _ { k } ( t ) | \boldsymbol { Q } ( t ) \bigg\} \\
& = \mathbb { E } \bigg\{ \sum _ { i \in \mathcal { S } } Q _ { i } ^ { S } ( t ) \!\!\!\!\!\!\!\!\!\!\!\!\!\!\sum _ { f \in \mathcal { F } ( A ( t ) ) : ( i , i ) \in f } \!\!\!\!\!\!\!\!\!\!\!\!\!\!\!\!\!\lambda _ { i } ( t ) q _ { f } ^ { \alpha } ( A ( t ) ) | \boldsymbol { Q } ( t ) \bigg\}\\
& + \mathbb { E } \bigg\{ \sum _ { j \in \mathcal { C } } \sum _ { i \in \mathcal { S } } \sum _ { f \in \mathcal { F } ( A ( t ) ) : ( i , j ) \in f } \!\!\!\!\!\!\!\!\!\!\!\!\!\!\!\!\!\lambda _ { i } ( t ) q _ { f } ^ { \alpha } ( A ( t ) ) Q _ { j } ^ { C } ( t ) | \boldsymbol { Q } ( t ) \bigg\}\\
& = \sum _ { Z \in \mathcal { P } ( \mathcal { N } ) } \sum _ { i \in \mathcal { S } } Q _ { i } ^ { S } ( t ) \!\!\!\!\!\!\!\!\!\sum _ { f \in \mathcal { F } ( Z ) : ( i , i ) \in f } \!\!\!\!\!\!\!\!\!\!\overline { \lambda } _ { i } \cdot q _ { f } ^ { \alpha } ( Z )\\
& + \mathbb { E } \bigg\{ \sum _ { j \in \mathcal { C } } Q _ { j } ^ { C } ( t ) \bigg( \sum _ { i \in \mathcal { S } } \sum _ { f \in \mathcal { F } ( A ( t ) ) : ( i , j ) \in f } \!\!\!\!\!\!\!\!\!\!\!\!\!\!\!\lambda _ { i } ( t ) q _ { f } ^ { \alpha } ( A ( t ) ) \bigg) | \boldsymbol { Q } ( t ) \bigg\}\\
& = \sum _ { Z \in \mathcal { P } ( \mathcal { N } ) } \sum _ { i \in \mathcal { S } } Q _ { i } ^ { S } ( t ) \sum _ { f \in \mathcal { F } ( Z ) : ( i , i ) \in f } \overline { \lambda } _ { i } \cdot q _ { f } ^ { \alpha } ( Z )\\
& + \mathbb { E } \bigg\{ \sum _ { j \in \mathcal { C } } Q _ { j } ^ { C } ( t ) \!\!\bigg( \sum _ { i \in \mathcal { S } } \sum _ { f \in \mathcal { F } ( A ( t ) ) } \!\!\!\!\!\!\!\!\mathds { 1 } _ { ( i , j ) \in f } \lambda _ { i } ( t ) q _ { f } ^ { \alpha } ( A ( t ) )\!\! \bigg) \!\!| \boldsymbol { Q } ( t ) \!\!\bigg\}\\
& = \sum _ { Z \in \mathcal { P } ( \mathcal { N } ) } p ( Z ) \sum _ { i \in \mathcal { S } } Q _ { i } ^ { S } ( t ) \sum _ { f \in \mathcal { F } ( Z ) : ( i , i ) \in f } \overline { \lambda } _ { i } \cdot q _ { f } ^ { \alpha } ( Z ) \\
& + \mathbb { E } \bigg\{ \sum _ { j \in \mathcal { C } } Q _ { j } ^ { C } ( t ) \bigg( \sum _ { f \in \mathcal { F } ( A ( t ) ) } \!\!\!\!\!\!q _ { f } ^ { \alpha } ( A ( t ) )\!\!\!\!\!\! \sum _ { i : ( i , j ) \in f }\!\!\!\!\!\! \lambda _ { i } ( t ) \bigg) | \boldsymbol { Q } ( t ) \bigg\}\\
& = \sum _ { i \in \mathcal { S } } Q _ { i } ^ { S } ( t ) \!\!\!\!\sum _ { Z \in \mathcal { P } ( \mathcal { N } ) } p ( Z ) \!\!\!\!\!\!\sum _ { f \in \mathcal { F } ( Z ) : ( i , i ) \in f }  \!\!\!\!\!\!\!\!q _ { f } ^ { \alpha } ( Z ) \cdot \overline { \lambda } _ { i } \\
& + \sum _ { j \in C } Q _ { j } ^ { C } ( t ) \!\!\!\!\!\!\sum _ { Z \in \mathcal { P } ( \mathcal { N } ) } \!\!\!\!\!p ( Z ) \!\bigg( \sum _ { f \in \mathcal { F } ( A ( t ) ) } \!\!\!\!\!\!\!\!q _ { f } ^ { \alpha } ( A ( t ) )\!\! \sum _ { i ( i , j ) \in f } \overline { \lambda } _ { i } \bigg).
\end{aligned}
\end{equation}

By applying (\ref{E-3}), (\ref{E-2}), and (\ref{E-1}), we have
\begin{equation}
\begin{aligned}
& \mathbb { E } [ L ( \boldsymbol { Q } ( t + 1 ) ) - L ( \boldsymbol { Q } ( t ) ) | \boldsymbol { Q } ( t ) ]\\
& \leq B - \sum _ { i \in \mathcal { S } } Q _ { i } ^ { S } ( t ) \cdot \overline { \mu } _ { i } - \sum _ { j \in \mathcal { C } } Q _ { j } ^ { C } ( t ) \cdot \overline { \mu } _ { j } - V  | \mathcal { S } | \lambda _ { \text{max} }  x _ { \text{min} }\\
& + \sum _ { i \in \mathcal { S } } Q _ { i } ^ { S } ( t ) \sum _ { Z \in \mathcal { P } ( \mathcal { N } ) } p ( Z ) \sum _ { f \in \mathcal { F } ( Z ) : ( i , i ) \in f } q _ { f } ^ { \alpha } ( Z ) \cdot \overline { \lambda } _ { i }\\
& + \sum _ { j \in \mathcal { C } } Q _ { j } ^ { C } ( t ) \!\!\!\!\!\sum _ { Z \in \mathcal { P } ( \mathcal { N } ) } \!\!\!\!p ( Z ) \bigg( \sum _ { f \in \mathcal { F } ( A ( t ) ) } \!\!\!\!\!q _ { f } ^ { \alpha } ( A ( t ) ) \sum _ { i : ( i , j ) \in f } \overline { \lambda } _ { i } \bigg)\\
& = B - V  | \mathcal { S } |  \lambda _ { \text{max} }  x _ { \text{min} }\\
& + \sum _ { i \in \mathcal { S } } Q _ { i } ^ { S } ( t ) \bigg[ \bigg( \sum _ { Z \in \mathcal { P } ( \mathcal { N } ) } \!\!\!\!\!p ( Z ) \!\!\!\!\!\!\!\sum _ { f \in \mathcal { F } ( Z ) : ( i , i ) \in f } \!\!\!\!\!\!\!\!\!q _ { f } ^ { \alpha } ( Z ) \cdot \overline { \lambda } _ { i } \bigg) - \overline { \mu } _ { i } \bigg]\\
\nonumber
\end{aligned}
\end{equation}
\begin{equation}
\begin{aligned}
& + \sum _ { j \in \mathcal { C } } Q _ { j } ^ { C } ( t ) \bigg[ \bigg( \sum _ { Z \in \mathcal { P } ( \mathcal { N } ) } \!\!\!\!\!p ( Z ) \!\!\!\!\!\!\!\sum _ { f \in \mathcal { F } ( A ( t ) ) } \!\!\!\!\!\!\!\!\!q _ { f } ^ { \alpha } ( A ( t ) )\!\!\!\! \sum _ { i : ( i , j ) \in f } \!\!\!\!\!\!\overline { \lambda } _ { i } \bigg) \!- \!\overline { \mu } _ { j } \bigg].
\end{aligned}
\end{equation}
According to (\ref{instance}), we have
\begin{equation}
\begin{array}{l}
\displaystyle
\mathbb { E } [ L ( \boldsymbol { Q } ( t + 1 ) ) - L ( \boldsymbol { Q } ( t ) ) | \boldsymbol { Q } ( t ) ]\\
\displaystyle
\leq \hat { B } - \epsilon \sum _ { i \in \mathcal { S } } Q _ { i } ^ { S } ( t ) - \epsilon \sum _ { j \in \mathcal { C } } Q _ { j } ^ { C } ( t )\\
\displaystyle
= \hat { B } - \epsilon \bigg( \sum _ { i \in \mathcal { S } } Q _ { i } ^ { S } ( t ) + \sum _ { j \in \mathcal { C } } Q _ { j } ^ { C } ( t ) \bigg),
\end{array}
\end{equation}
where $\hat { B }$ is defined as 
\begin{equation}\label{definition: b_hat}
\begin{aligned}
\hat { B }(V) & \triangleq B - V  | \mathcal { S } |  \lambda _ { \text{max}} x _ { \text{min} } \\
&= B + V  | \mathcal { S } | \lambda _ { \text{max} }  \max \bigg\{ w _ { \text{max} } , m _ { \text{max} } \bigg\}.	
\end{aligned}
\end{equation}

By summing $\mathbb { E } [ L ( \boldsymbol { Q } ( t + 1 ) ) - L ( \boldsymbol { Q } ( t ) ) | \boldsymbol { Q } ( t ) ]$ over $t = 0, \ldots,T-1$ and taking expectation on both sides, we have

\begin{equation}
\begin{aligned}
&\mathbb { E } [ L ( \boldsymbol { Q } ( T ) ) - L ( \boldsymbol { Q } ( 0 ) ]\\
& \leq T \cdot \hat { B }(V) - \epsilon \sum _ { t = 0 } ^ { T - 1 } \mathbb { E } \bigg\{ \sum _ { i \in \mathcal { S } } Q _ { i } ^ { S } ( t ) + \sum _ { j \in \mathcal { C } } Q _ { j } ^ { C } ( t ) \bigg\}.
\end{aligned}
\end{equation}

Dividing by $T$ at both sides, we can obtain
\begin{equation}
\begin{aligned}
& \frac { \epsilon } { T } \sum _ { t = 0 } ^ { T - 1 } \mathbb { E } \bigg\{ \sum _ { i \in \mathcal { S } } Q _ { i } ^ { S } ( t ) + \sum _ { j \in \mathcal { C } } Q _ { j } ^ { C } ( t ) \bigg\}\\
& \leq \hat { B }(V) - \frac { \mathbb { E } [ L ( \boldsymbol { Q } ( T ) ) - L ( \boldsymbol { Q } ( 0 ) ] } { T }.
\end{aligned}
\end{equation}

Since $\boldsymbol { Q } ( T ) \geq 0$  and $\boldsymbol { Q } ( 0 ) \geq 0$, then taking lim-sup as $T \rightarrow \infty$ and with the positiveness of $\epsilon$, we have 

\begin{equation}
\limsup _ { T \rightarrow \infty } \frac { 1 } { T } \sum _ { t = 0 } ^ { T - 1 } \mathbb { E } \bigg\{ \sum _ { i \in \mathcal { S } } Q _ { i } ^ { S } ( t ) + \sum _ { j \in \mathcal { C } } Q _ { j } ^ { C } ( t ) \bigg\} \leq \frac { \hat { B }(V) } { \epsilon }.
\end{equation}
Recalling the definition of $\hat{B}(V)$ in (\ref{definition: b_hat}), we have
\begin{equation}
\begin{aligned}
&\limsup _ { T \rightarrow \infty } \frac { 1 } { T } \sum _ { t = 0 } ^ { T - 1 } \mathbb { E } \bigg\{ \sum _ { i \in \mathcal { S } } Q _ { i } ^ { S } ( t ) + \sum _ { j \in \mathcal { C } } Q _ { j } ^ { C } ( t ) \bigg\} \\
& \leq \frac { B } { \epsilon } + V \cdot | \mathcal { S } | \cdot \lambda _ { \max } \cdot \max \bigg\{ w _ { \max } , m _ { \max } \bigg\}.
\end{aligned}
\end{equation}
\IEEEQED

\section{Proof of Theorem \ref{theorem2}}  \label{proof of theorem 2}
Consider an optimal policy $\alpha ^ { * }$ and its corresponding probability distributions $\mathbf { q } ^ { * } = \bigg[ q _ { Z } ^ { * } ( f ) , \forall Z \in \mathcal { P } ( \mathcal { N } ) , \forall f \in \mathcal { F } ( Z ) \bigg].$ Let $f ^ {*} ( t )$ be the super arm set selected by policy $\alpha ^ { * }$ in round $t$. Matrix $\mathbf { I } ^ { * } ( t ) \triangleq (I_{i,j})_{i \in S, j \in \{i\} \cup C}$ is the associated action matrix. Accordingly, we have 
\begin{equation}\label{definition: optimal regret}
	\begin{aligned}
		R ^ { * } 
		& =\!\!\!\!  \sum _ { Z \in \mathcal { P } ( N ) } p ( Z ) \!\!\! \sum _ { f \in \mathcal { F } ( Z ) } q _ { Z } ^ { * } ( f ) \sum _ { n \in f } \overline { x } _ { n }
		& = \mathbb { E } \bigg[ \sum _ { ( i , j ) \in f ^ { * } ( t ) } \overline { x } _ { i , j } \bigg].
	\end{aligned}
\end{equation}

By (\ref{definition: optimal regret}) and (\ref{regret}), we can rewrite the regret of LASAC as
\begin{equation}
\begin{aligned}
R _ { \mathrm { \text{LASAC} } } ( T ) &= R ^ { * } - \mathbb { E } \bigg\{ \frac { 1 } { T } \sum _ { t = 1 } ^ { T } \sum _ { n \in f ( t ) } X _ { n } ( t ) \bigg\} \\
& = \frac { 1 } { T } \sum _ { t = 1 } ^ { T } \bigg\{ R ^ { * } -  \mathbb { E } \bigg[ \sum _ { ( i , j ) \in f ( t ) } \overline x _ { i , j } ( t ) \bigg] \bigg\} \\
& = \frac { 1 } { T } \sum _ { t = 1 } ^ { T }  \mathbb { E } \bigg[  \underbrace {\sum _ { ( i , j ) \in f^{*} ( t ) } \overline x _ { i , j } ( t ) - \sum _ { ( i , j ) \in f ( t ) } \overline x _ { i , j } ( t ) }_ { \Delta R ( t ) }\bigg].\\
\end{aligned}
\end{equation}
Next, we define $\Delta R ( t )$ as the difference between the expected rewards which are achieved by the optimal policy $\alpha ^ { * }$ and LASAC during time slot $t$. In particular,
\begin{equation}
\begin{array}{rl}
\Delta R ( t ) & \displaystyle
\triangleq  \sum _ { ( i , j ) \in f^{*} ( t ) } \overline x _ { i , j } ( t ) - \sum _ { ( i , j ) \in f ( t ) } \overline x _ { i , j } ( t ) \\
& = \displaystyle
\sum _ {i \in \mathbf { S }}\bigg[I _ { i , i }^ { * } ( t ) \overline { x } _ { i , i }  + \sum _ {j \in \mathcal{ C }}I _ { i , j }^ { * } ( t ) \overline { x } _ { i , j } \bigg] \\
& - \displaystyle
\sum _ {i \in \mathbf { S }}\bigg[I _ { i , i } ( t ) \overline { x } _ { i , i }  + \sum _ {j \in \mathcal{ C }}I _ { i , j } ( t ) \overline { x } _ { i , j } \bigg].
\end{array}
\end{equation}
By adding $\Delta R ( t )$ (scaled by $V \lambda_{\text{max}}$) to  (\ref{definition:delta-t}), we have 
\begin{equation} \label{definition:drift-plus-regret}
\begin{aligned}
\nonumber
& L ( \mathbf { Q } ( t + 1 ) ) - L ( \mathbf { Q } ( t ) ) + V\lambda_{\text{max}}\Delta R ( t )\\
& \stackrel{(a)}\leq B - \sum _ { i \in \mathcal { S } } Q _ { i } ^ { S } ( t ) \mu _ { i } ^ { S } ( t ) - \sum _ { j \in \mathcal { C } } Q _ { j } ^ { C } ( t ) \mu _ { j } ^ { C } ( t ) \\
&+ \sum _ { i \in \mathcal { S } } \lambda _ { \text{max} } \bigg[ Q _ { i } ^ { S } ( t ) I _ { i , i } ( t ) + \sum _ { j \in \mathcal { C } } Q _ { j } ^ { C } ( t ) I _ { i , j } ( t ) \bigg]\\
& + V\lambda_{\text{max}} \sum _ {i \in \mathbf { S }}\bigg[I _ { i , i }^ { * } ( t ) \overline { x } _ { i , i }  + \sum _ {j \in \mathcal{ C }}I _ { i , j }^ { * } ( t ) \overline { x } _ { i , j } \bigg] \\
\end{aligned}
\end{equation}
\begin{equation}
\begin{aligned}
& - V\lambda_{\text{max}}\sum _ {i \in \mathbf { S }}\bigg[I _ { i , i } ( t ) \overline { x } _ { i , i }  + \sum _ {j \in \mathcal{ C }}I _ { i , j } ( t ) \overline { x } _ { i , j } \bigg]\\
& = B + \sum _ { i \in \mathcal { S } }  [\lambda _ { \text{max} } Q _ { i } ^ { S } ( t ) - V\lambda_{\text{max}}\overline { x } _ { i , i }   ](I _ { i , i } ( t ) - I ^ { *} _ {i , i} ( t ) )  \\
& + \sum _ { i \in \mathcal { S } }  [ \sum _ {j \in \mathcal{ C }} \lambda _ { \text{max} } Q _ { j } ^ { C } ( t )  - V\lambda_{\text{max}}\overline { x } _ { i , j }  ] (I _ { i , j } ( t ) - I ^ { *} _ {i , j} ( t ) ) \\
& + \sum _ { i \in \mathcal { S } } Q _ { i } ^ { S } ( t ) ( \lambda _ { \text{max} }  I ^ { *} _ {i , i} ( t ) - \mu _ { i } ^ { S } ( t )) \\
&+ \sum _ {j \in \mathcal{ C }} Q _ { j } ^ { C } ( t )(\sum _ { i \in \mathcal { S } }  \lambda _ { \text{max} } I ^ { *} _ {i , j} ( t ) - \mu _ { j } ^ { C } ( t ) )\\
& \stackrel{(b)}\leq B + \sum _ { i \in \mathcal { S } }  [\lambda _ { \text{max} } Q _ { i } ^ { S } ( t ) - V\lambda_{\text{max}}\overline { x } _ { i , i }   ](I _ { i , i } ( t ) - I ^ { *} _ {i , i} ( t ) )  \\
& + \sum _ { i \in \mathcal { S } }  [ \sum _ {j \in \mathcal{ C }} \lambda _ { \text{max} } Q _ { j } ^ { C } ( t )  - V \lambda _{\text{max}}\overline { x } _ { i , j }  ] (I _ { i , j } ( t ) - I ^ { *} _ {i , j} ( t ) ) \\
& = B + \!\! \lambda_{\text{max}}\sum _ { i \in \mathcal { S } }  \big[ \!\!\! \sum _ {k \in \{ i \} \cup \mathcal{ C }} (Q_{k} ( t )  - V \overline { x } _ { i , k }  )(I _ { i , k } ( t ) - I ^ { *} _ {i , k} ( t ) ) \big],\\
\end{aligned}
\end{equation}
where $(a)$ is due to (\ref{definition:delta-t}), $\lambda_{i,k}(t)$ is upper bounded by $\lambda_{\text{max}}$ and $(b)$ is due to that service capacities of switches and servers service  are greater than the number of requests. 

By defining $Q_{k}(t) \triangleq Q^{s}_i(t)$ if $k = i$ and $Q_{k}(t) \triangleq Q^{c}_j(t)$ otherwise, 
then taking the expectation of (\ref{definition:drift-plus-regret}), we obtain
\begin{equation}\label{definition:expectation-drift-plus-regret}
\begin{aligned}
\nonumber
& \mathbb { E } [ L ( \mathbf { Q } ( t + 1 ) ) - L ( \mathbf { Q } ( t ) ) + V\lambda_{\text{max}}\Delta R ( t ) ] \\
& \stackrel{a}\leq B  + \\
&\lambda_{\text{max}} \mathbb { E } \bigg\{\underbrace{\sum _ { i \in \mathcal { S } }  \bigg[\sum _ {k \in \{ i \} \cup \mathcal{ C }} (Q_{k} ( t )  - V \overline { x } _ { i , k }  )(I _ { i , k } ( t ) - I ^ { *} _ {i , k} ( t ) ) \bigg]}_{Z_1(t)}\bigg\} .\\
\end{aligned}
\end{equation}
Next, we define $Z_1(t)$ as follows:
\begin{equation}\label{definition:Z-1-t}
\begin{aligned}
Z_1(t) \triangleq \sum _ { i \in \mathcal { S } }  \bigg[\sum _  {k \in \{ i \} \cup \mathcal{ C }} (Q_{k} ( t )  - V \overline { x } _ { i , k }  )(I _ { i , k } ( t ) - I ^ { *} _ {i , k} ( t ) ) \bigg].
\end{aligned}
\end{equation} 
Then summing (\ref{definition:expectation-drift-plus-regret}) over $t \in \{ 0, \dots, T - 1 \}$, dividing both sides of the inequality by $T V \lambda_{\text{max}}$, we have
\begin{equation}
\begin{aligned}
& \frac { 1 } { T V\lambda_{\text{max}}} \mathbb { E } [ L ( \mathbf { Q } ( T ) ) - L ( \mathbf { Q } ( 0 ) ) ] + \frac { 1 } { T } \sum _ { t = 0 } ^ { T - 1 } \mathbb { E } [ \Delta R ( t ) ]\\
& \leq \frac { B } { V\lambda_{\text{max}}} + \frac { 1 }{ T V }\sum _ { t = 0 } ^ { T - 1 } \mathbb { E } \bigg[ Z _ { 1 } ( t ) \bigg].
\end{aligned}
\end{equation}
Since $L ( \mathbf { Q } ( T ) ) \geq 0$ and $L ( \mathbf { Q } ( 0 ) ) = 0$, then
\begin{equation}\label{inequation: upper_bound_of_average_regret}
\begin{aligned}
\frac { 1 } { T } \sum _ { t = 0 } ^ { T - 1 } \mathbb { E } [ \Delta R ( t ) ] \leq \frac { B } { V\lambda_{max}} + \frac { 1 }{ T V }\sum _ { t = 0 } ^ { T - 1 } \mathbb { E } \bigg[ Z _ { 1 } ( t ) \bigg].
\end{aligned}
\end{equation}

Next, we need to find an upper bound for $\mathbb { E } [ Z _ { 1 } ( t ) ]$. Consider a policy $\pi ^ { \prime }$, in each time slot $t$, choose a super arm $f'(t)$ as follows
\begin{equation}\label{arg}
\begin{aligned}
f'(t)\in \arg\min\sum_{(i, k)\in f} (Q_{k}  ( t )- V \cdot \overline { x } _ { i , k } ( t )).
\end{aligned}
\end{equation}
Therefore, we have
\begin{equation}\label{relationship}
\sum_{(i, k)\in f(t)} (Q_{k}  ( t )- V \cdot \tilde { x } _ { i , k } ( t )) \leq \sum_{(i, k)\in f'(t)} (Q_{k}  ( t )- V \cdot \tilde { x } _ { i , k } ( t )),
\end{equation}
thus we reach an upper bound on $Z_1(t)$, \textit{i.e.},
\begin{equation}
\begin{aligned}
Z _ { 1 } ( t ) &= \sum _ { i \in \mathcal { S } }  \bigg[\sum _ {k \in \{ i \} \cup \mathcal{ C }} (Q_{k} ( t )  - V \overline { x } _ { i , k }  )(I _ { i , k } ( t ) - I ^ { *} _ {i , k} ( t ) ) \bigg]\\
& = \!\!\!\!\!\!\sum _ { (i,k) \in f(t)}\!\!\!\!(Q_{k} ( t ) - V\ \overline { x } _ { i , k } )-\!\!\!\!\!\!\!\sum _ { (i,k) \in f^{*}(t) } \!\!\!\!(Q_{k} ( t ) - V  \overline { x } _ { i , k })\\
& \stackrel{(a)} \leq \!\!\!\!\!\!\sum _ { (i,k) \in f(t)}\!\!\!\!(Q_{k} ( t ) - V\ \overline { x } _ { i , k } )-\!\!\!\!\!\!\! \sum _ { (i,k) \in f^{'}(t) } \!\!\!\!(Q_{k}  ( t ) - V  \overline { x } _ { i , k })\\
& \stackrel{(b)} \leq \!\!\!\!\!\!\sum _ { (i,k) \in f(t)}\!\!\!\!(Q_{k} ( t ) - V\ \overline { x } _ { i , k } )-\!\!\!\!\!\!\! \sum _ { (i,k) \in f^{'}(t) } \!\!\!\!(Q_{k}  ( t ) - V  \overline { x } _ { i , k })\\
& + \!\!\!\!\!\!\sum _ { ( i , k ) \in f ^ { \prime } ( t ) } \!\!\!\!\bigg( Q_{k}  ( t ) - V \tilde { x } _ { i , k } ( t ) \bigg) -\!\!\!\!\!\!\! \sum _ { ( i , k ) \in f ( t ) } \!\!\!\!\bigg( Q_{k}  ( t ) - V  \tilde { x } _ { i , k } ( t ) \bigg)\\
& = V \Bigg[ \!\!\underbrace{\sum _ { ( i , k ) \in f ( t ) } \!\!\!\!\!\!\!\bigg(\tilde { x } _ { i , k } ( t ) - \overline { x } _ { i , k } \bigg)  }_{Z_2(t)} + \!\!\!\!\!\underbrace { \sum _ { ( i , k ) \in f ^ { \prime } ( t ) } \!\!\!\!\!\!\!\bigg( \overline { x } _ { i , k } - \tilde { x } _ { i , k } ( t ) \bigg) } _ { Z _ { 3 } ( t )}\Bigg],
\end{aligned}
\end{equation}

where $(a)$ is from (\ref{arg}) and $(b)$ is from (\ref{relationship}) and we define $Z_2(t)$ and $Z_3(t)$ as 
\begin{equation}
Z_2(t) \triangleq \sum _ { ( i , k ) \in f ( t ) } \bigg( \tilde { x } _ { i , k } ( t ) - \overline { x } _ { i , k } \bigg) \\
\end{equation}
and 
\begin{equation}
Z_3(t)\triangleq \sum _ { ( i , k ) \in f ^ { \prime } ( t ) } \bigg( \overline { x } _ { i , k } - \tilde { x } _ { i , k } ( t ) \bigg).
\end{equation}

In order to bound $Z_2(t)$, we consider an arbitrary arm $( i , k )_{i \in \mathcal{ S },k \in \{ i \} \cup \mathcal{ C }}$ and in time slot $t$. 
Let $t _ { a } ^ { i , k }$ be the time slot during which arm $( i , k )$ is played for the $a$-th time. 
We know that $h _ { i , k } ( t ) \triangleq \sum _ { \tau = 1 } ^ { t } I _ { i , k } ( \tau )$ which represents the number of arm $(i,k)$ which has been played by the end of round $t$. Obviously, we have $I _ { i , k } ( t _ { a } ^ { i , k } ) = 1 , h _ { i , k } ( t _ { a } ^ { i , k } ) = a$ and $h _ { i , k } ( t _ { a } ^ { i , k } - 1) = a - 1$ for all $a \in \{ 1,2 , \ldots , h _ { i , k } ( T - 1 ) \}$. Accordingly, we also have 
\begin{equation}
0 \leq t _ { 1 } ^ { i , k } < t _ { 2 } ^ { i , k } < \cdots < t _ { h _ { i , k } ( T - 1 ) } ^ { i , k } < T.
\end{equation}
By defining
\begin{equation}
\Delta X _ { i , k } ( t ) \triangleq \bigg\{ \tilde { x } _ { i , k } ( t ) - \overline { x } _ { i , k } > 0 \bigg\},
\end{equation}
and $U ^ { c }$ as the complement event of $U$, and $\mathds{1}_{\{ \cdot\}}$ as the indicator function. The expectation of $Z_2(t)$ is bounded by

\begin{equation}\label{expectation:Z-2-t}
\begin{aligned}
\nonumber
\mathbb { E } \bigg[ Z _ { 2 } ( t ) \bigg] & = \sum _ { ( i , k ) \in f ( t ) } \bigg( \tilde { x } _ { i , k } ( t ) - \overline { x } _ { i , k } \bigg)\\
\end{aligned}
\end{equation}
\begin{equation}
\begin{aligned}
& = \mathbb { E } \bigg[\sum _ { i \in \mathcal{ S } } \sum _ { k \in \{ i \} \cup \mathcal{ C }} \bigg( \tilde { x } _ { i , k } ( t ) - \overline { x } _ { i , k } \bigg)I_{ i , k}(t)\bigg]\\
& = \sum _ { i \in \mathcal{ S } } \sum _ { k \in \{ i \} \cup \mathcal{ C }}\mathbb { E } \bigg[ \bigg( \tilde { x } _ { i , k } ( t ) - \overline { x } _ { i , k } \bigg)I_{ i , k}(t) \mathds{1}_{\Delta X _ { i , k } ( t )}\bigg] \\
& + \sum _ { i \in \mathcal{ S } } \sum _ { k \in \{ i \} \cup \mathcal{ C }}\mathbb { E } \bigg[ \bigg( \tilde { x } _ { i , k } ( t ) - \overline { x } _ { i , k } \bigg)I_{ i , k}(t) \mathds{1}_{\Delta X _ { i , k } ^ {c}( t )}\bigg]\\
& \stackrel{(a)} \leq \sum _ { i \in \mathcal{ S } } \sum _ { k \in \{ i \} \cup \mathcal{ C }} \!\!\! \mathbb { E }\bigg[ \underbrace{\bigg( \tilde { x } _ { i , k } ( t ) - \overline { x } _ { i , k } \bigg)I_{ i , k}(t) \mathds{1}_{\Delta X _ { i , k } ( t )}}_{J_1(t)}\bigg],
\end{aligned}
\end{equation}
where $(a)$ is due to $\tilde { x } _ { i , k } ( t ) - \overline { x } _ { i , k } \leq 0 $, 
provided that event $\Delta X_{i,k}^{c}(t)$ occurs.
Next, by defining $J _ { 1 } ( t ) \triangleq \big( \tilde { x } _ { i , k } ( t ) - \overline { x } _ { i , k } \big)I_{ i , k}(t) \mathds{1}_{\Delta X _ { i , k } ( t )}$ and another event 
\begin{equation}
F _ { i , k } ( t ) \triangleq \bigg\{ \hat { x } _ { i , k } ( t - 1 ) - \overline { x } _ { i , k } \leq \beta \sqrt { \frac { \log t } { 4 h _ { i , k } ( t - 1 ) } } \bigg\},
\end{equation}
and summing $J _ { 1 } ( t )$ over $t \in \{ 0 , \ldots , T - 1 \}$,
we have
\begin{equation}\label{definition:J-1-t}
\begin{aligned}
& \sum _ { t = 0 } ^ { T - 1 } J _ { 1 } ( t )  \stackrel{(a)} =  \sum _ { a = 1 } ^ { h _ { i , k } ( T - 1 ) } \bigg( \tilde { x } _ { i , k } ( t _ { a } ^ { i , k } ) - \overline { x } _ { i , k } \bigg)  \mathds{1} _ { \Delta X _ { i , k } ( t _ { a } ^ { i , k } ) }\\
& \stackrel{(b)} \leq -x_{\text{min}} + \sum _ { a = 2 } ^ { h _ { i , k } ( T - 1 ) } \bigg( \tilde { x } _ { i , k } ( t _ { a } ^ { i , k } ) - \overline { x } _ { i , k } \bigg)  \mathds{1} _ { \Delta X _ { i , k } ( t _ { a } ^ { i , k } ) }\\ \nonumber
\end{aligned}
\end{equation}
\begin{equation}
\begin{aligned}
& = -x_{\text{min}} \\
& + \sum _ { a = 2 } ^ { h _ { i , k } ( T - 1 ) } \bigg[ \tilde { x } _ { i , k } ( t _ { a } ^ { i , k } ) - \overline { x } _ { i , k } \bigg]  \mathds{1} _ { \Delta X _ { i , k } ( t _ { a } ^ { i , k } ) }\mathds{1} _ { F _ { i , k } ( t _ { a } ^ { i , k } ) } \\
&+ \sum _ { a = 2 } ^ { h _ { i , k } ( T - 1 ) } \bigg[ \tilde { x } _ { i , k } ( t _ { a } ^ { i , k } ) - \overline { x } _ { i , k } \bigg]  \mathds{1} _ { \Delta X _ { i , k } ( t _ { a } ^ { i , k } ) } \mathds{1} _ { F _ { i , k } ^ { c }( t _ { a } ^ { i , k } ) }\\
& \stackrel{(c)} \leq  -x_{\text{min}} + 
\!\!\!\!\!
\sum _ { a = 2 } ^ { h _ { i , k } ( T - 1 ) } 
\!\!\!
\underbrace{\bigg[ \tilde { x } _ { i , k } ( t _ { a } ^ { i , k } ) - \overline { x } _ { i , k } \bigg]  \mathds{1} _ { \Delta X _ { i , k } ( t _ { a } ^ { i , k } ) \cap F _ { i , k } ( t _ { a } ^ { i , k } )} }_{J_2(t _ { a } ^ { i , k })}\\
& + \sum _ { a = 2 } ^ { h _ { i , k } ( T - 1 ) } \underbrace{-x_{\text{min}} \mathds{1} _ { F _ { i , k } ^ { c }( t _ { a } ^ { i , k } ) }}_{J_3(t _ { a } ^ { i , k })},
\end{aligned}
\end{equation}
where $x_{\text{min}} \triangleq -\max\{w_{\text{max}},m_{\text{max}}\}$, 
equality $(a)$ is due to $d _ { i , k} ( t _ { a } ^ { i , k } ) = 1$ for all $a \in \{ 1,2 , \ldots , h _ { i,k } ( T - 1 ) \}$ and $d_{i,k}(t) = 0$ for other $t$; 
inequality $(b)$ is due to $\tilde { x } _ { i , k } ( t _ { 1 } ^ { i , k } )- \overline { x } _ { i , k }  \leq -x_{\text{min}}$; 
inequality $(c)$ is due to $(\tilde { x } _ { i , k }( t _ { 1 } ^ { i , k }) - \overline { x } _ { i , k })  \mathds{1} _ { \Delta X _ { i , k } }\leq -x_{\text{min}}$.

Next, we define 
\begin{equation}
	J_2(t _ { a } ^ { i , k }) \triangleq \Big( \tilde { x } _ { i , k } ( t _ { a } ^ { i , k } ) - \overline { x } _ { i , k } \Big)  \mathds{1} _ { \Delta X _ { i , k } ( t _ { a } ^ { i , k } ) \cap F _ { i , k } ( t _ { a } ^ { i , k } )}
\end{equation}
and
\begin{equation}
	J_3(t _ { a } ^ { i , k }) \triangleq -x_{\text{min}} \mathds{1} _ { F _ { i , k } ^ { c }( t _ { a } ^ { i , k } ) }.
\end{equation}
We want to give upper bounds for $\sum _ { a = 2 } ^ { h _ { i , k } ( T - 1 ) }\mathbb { E } \big[ J_2(t _ { a } ^ { i , k }) \big]$ and $\sum _ { a = 2 } ^ { h _ { i , k } ( T - 1 ) }\mathbb { E } \big[ J_3(t _ { a } ^ { i , k }) \big]$.

First, to bound $\sum _ { a = 2 } ^ { h _ { i , k } ( T - 1 ) }\mathbb { E } \big[ J_2(t _ { a } ^ { i , k }) \big]$, we consider $t _ { a } ^ { i , k }$ for all $a \in \big\{ 2 , \ldots , h _ { i , k } ( T - 1 ) \big\}$. 
Suppose that event $F _ { i , k } ( t _ { a } ^ { i , k } )$ happens.
Then we have 
\begin{equation}\label{inequation: mean-reward}
\hat { x } _ { i , k } ( t _ { a } ^ { i , k } - 1 ) - \overline x _ { i , k } \leq \beta \sqrt { \frac { \log t _ { a } ^ { i , k } } { 4 h _ { i , k } \big( t _ { a } ^ { i , k } - 1 \big) } }.\\
\end{equation}
From (\ref{equation: update ucb}) and (\ref{definition: ucb_bound}), we also have 
\begin{equation}\label{inequation: usb-reward}
\tilde x _ { i , k } (t _ { a } ^ { i , k }) \leq \hat { x } _ { i , k } \big( t _ { a } ^ { i , k } - 1 \big)  + \beta \sqrt { \frac { \log t _ { a } ^ { i , k } } { 4 h _ { i , k } \big( t _ { a } ^ { i , k } - 1 \big) } }.\\
\end{equation}
Combining (\ref{inequation: mean-reward}) and (\ref{inequation: usb-reward}), it turns out
\begin{equation}
\tilde x _ { i , k } (t _ { a } ^ { i , k }) - \overline x _ { i , k }  \leq 2 \beta \sqrt { \frac { \log t _ { a } ^ { i , k } } { 4 h _ { i , k } \big( t _ { a } ^ { i , k } - 1 \big) } },\\
\end{equation}
which implies that for all $a \in \{2,\ldots,h_{i,k}(T-1)\}$, 
\begin{equation}\label{definition:J-2-t}
\begin{aligned}
J_2(t _ a ^ { i , k }) & = \big( \tilde { x } _ { i , k } ( t _ { a } ^ { i , k } ) - \overline { x } _ { i , k } \big)  \mathds{1} _ { \Delta X _ { i , k } ( t _ { a } ^ { i , k } ) \cap F _ { i , k } ( t _ { a } ^ { i , k } )} \\
& \leq 2 \beta \sqrt { \frac { \log t _ { a } ^ { i , k } } { 4 h _ { i , k } \big( t _ { a } ^ { i , k } - 1 \big) } }.
\end{aligned}
\end{equation}

Summing $J_2(t _ a ^ { i , k })$ over $a \in {2,\ldots,h_{i,k}(T-1)}$ yields
\begin{equation}
\begin{aligned}
\sum_{ a = 2 } ^ { h _ { i , k }( T - 1 )}J_2(t _ a ^ { i , k }) & \stackrel{(a)}\leq  \sum_{ a = 2 } ^ { h _ { i , k }( T - 1 )} 2\beta \sqrt { \frac { \log t _ { a } ^ { i , k } } { 4 h _ { i , k } ( t _ { a } ^ { i , k } - 1 ) } }\\
& \stackrel{(b)} \leq \beta \sqrt{\log T}\sum_{ a = 2 } ^ { h _ { i , k }( T - 1 )} \frac{1}{\sqrt{a-1}}\\
& \leq \beta \sqrt{\log T} \bigg( 1 + \int _ { 1 } ^ { h _ { i , k } ( T - 1 ) } \frac { 1 } { \sqrt { x } } d x \bigg) \\
& \leq 2 \beta \sqrt{h _ { i , k } ( T - 1 ) \log T },
\end{aligned}
\end{equation}
where $(a)$ is by applying (\ref{definition:J-2-t}), 
$(b)$ is due to $t_a^{i,k} \leq T$ and $h_{i,k}(t_{a}^{i,k}) = a - 1$ for all $a \in \{2,\ldots,h_{i,k}(T-1)\}$. We have 
\begin{equation}\label{expectation:J-2-t}
\sum_{ a = 2 } ^ { h _ { i , k }( T - 1 )}\mathbb { E } \bigg[ J _ { 2 } \Big( t _ { a } ^ { i , k } \Big) \bigg] \leq 2 \beta \sqrt{\log T}\mathbb { E }\bigg[ \sqrt{h _ { i , k } ( T - 1 ) }\bigg].
\end{equation}

Next, we switch to bounding $\sum _ { a = 2 } ^ { h _ { i , k } ( T - 1 ) } \mathbb { E } \big[ J _ { 3 } ( t _ { a } ^ { i , k } ) \big]$. By applying Chernoff-Hoeffding bound, we have
\begin{equation}
\begin{aligned}
& \mathbb { E } \bigg[ \mathds{ 1 } _ { \Big\{ F _ { i , k } ^ { c } ( t _ { a } ^ { i , k } ) \Big\} } \bigg] = \mathbb { P } \Big\{ F _ { i , k } ^ { c } ( t _ { a } ^ { i , k} ) \Big\}\\
=& \mathbb { P } \!\Bigg\{ \hat { x } _ { i , k } ( t _ { a } ^ { i , k } - 1 ) - \overline { x } _ { i , k } 
> \beta \sqrt { \frac { \log t _ { a } ^ { i , k } } { 4 h _ { i , k } ( t _ { a } ^ { i , k } - 1) }  } \bigg\}\\
\leq & \frac { 1 }{{t _ { a } ^ { i , k }}^{\frac{\beta^2}{2}} }.
\end{aligned}
\end{equation}

Suppose that $\bigg\{ t _ { 2 } ^ { i } , \ldots , t _ { h _ { i } ( T - 1 ) } ^ { i } \bigg\} \subseteq \{ 1,2 , \ldots \}$, and that $\frac { \beta ^ { 2 } } { 2 } > 1$, \textit{i.e.}, $\beta > \sqrt{2}$, 
we have
\begin{equation}\label{expectation:F-c-t}
\begin{aligned}
\sum _ { a = 2 } ^ { h _ { i , k} ( T - 1 ) } \mathbb { E } \bigg[ \mathds { 1 } _ { \Big\{ F _ { i , k } ^ { c } ( t _ { a } ^ { i , k } ) \Big\} } \bigg] & \leq \sum _ { a = 2 } ^ { h _ { i , k} ( T - 1 ) } \frac { 1 }{{t _ { a } ^ { i , k }}^{\frac{\beta^2}{2}} } \\
& \leq \underbrace{\sum _ { t = 1 } ^ { \infty } \frac { 1 } { t ^ { \frac{\beta^2}{2} } }}_{G_{\beta}}.\\
\end{aligned}
\end{equation}
We define $G_{\beta} \triangleq \sum _ { t = 1 } ^ { \infty }  t ^ { -\frac { \beta ^ { 2 } } { 2 } } $ and we know that when $\frac { \beta ^ { 2 } } { 2 } > 1 $, $G$ has the property of convergence. Therefore, 
\begin{equation}\label{expectation:J-3-t}
\begin{aligned}
\sum _ { a = 2 } ^ { h _ { i , k } ( T - 1 ) } \mathbb { E } \bigg[ J _ { 3 } \bigg( t _ { a } ^ { i , k } \bigg) \bigg] \leq -x_{\text{min}}G_{\beta}.
\end{aligned}
\end{equation}
Taking expectation at both sides of (\ref{definition:J-1-t}), then combining (\ref{expectation:J-2-t}) and (\ref{expectation:J-3-t}), we have
\begin{equation}\label{expectation:J-1-t}
\begin{aligned}
\sum _ { t = 0 } ^ { T - 1 } \mathbb { E } \bigg[ J _ { 1 } ( t ) \bigg] &\leq -(G_{\beta}+1)x_{\text{min}} + \\
& 2 \beta \sqrt{\log T}\mathbb { E }\bigg[ \sqrt{h _ { i , k } ( T - 1 ) }\bigg].
\end{aligned}
\end{equation}

By summing (\ref{expectation:Z-2-t}) over $t \in \{ 0, \ldots, T-1\}$ and plugging (\ref{expectation:J-1-t}) thereinto, we have
\begin{equation}
\begin{array}{l}
\displaystyle
\sum_ {t = 0} ^{ T - 1 }\mathbb { E } \Big[ Z _ { 2 } ( t ) \Big]  
\leq  \sum _ { i \in \mathcal{ S } } \sum_{k \in \{ i \} \cup \mathcal{ C }}-(G_{\beta}+1)x_{\text{min}} \\
\displaystyle
+ \sum _ { i \in \mathcal{ S } } \sum_{k \in \{ i \} \cup \mathcal{ C }}\bigg(2 \beta \sqrt{\log T}\mathbb { E }\bigg[ \sqrt{h _ { i , k } ( T - 1 ) }\bigg]\bigg)\\
\displaystyle
\stackrel{(a)} \leq |S|(|C|+1)\bigg(2\beta \sqrt{T \log T} - (G_{\beta}+1)x_{\text{min}}\bigg).
\end{array}
\end{equation}

Next, we switch to bounding $Z_3(t)$. 
Consider an arbitrary arm $( i , k )_{i \in \mathcal{ S },k \in \{ i \} \cup \mathcal{ C }}$ and arbitrary time slot $t = 0,1 , \ldots , T - 1$. Recall that $Z_3(t) = \sum _ { ( i , k ) \in f ^ { \prime } ( t ) } \Big( \overline { x } _ { i , k } - \tilde { x } _ { i , k } ( t ) \Big)$. Let matrix $\mathbf { I } ^ { ' } ( t ) \triangleq \Big\{ I ^{ ' }_ { i , k } \Big\} _ { i \in S , k \in \{ i \} \cup C }$ be the action matrix corresponding to $f'(t)$. Also, recall that $\Delta X _ { i , k } ( t ) \triangleq \Big\{ \tilde { x } _ { i , k } ( t ) - \overline { x } _ { i , k } > 0 \Big\}.$ Similar to the derivation for $Z_2(t)$ in (\ref{expectation:Z-2-t}), we bound the expectation of $Z_3(t)$ as follows
\begin{equation}\label{expectation:Z-3-t}
\begin{aligned}
\mathbb { E } \Big[ Z _ { 3 } ( t ) \Big] \!\!\! =& \sum _ { ( i , k ) \in f ( t ) } \bigg( \overline { x } _ { i , k } - \tilde { x } _ { i , k } ( t ) \bigg)\\
=& \mathbb { E } \bigg[\sum _ { i \in \mathcal{ S } } \sum _ { k \in \{ i \} \cup \mathcal{ C }} \bigg(  \overline { x } _ { i , k } - \tilde { x } _ { i , k } ( t ) \bigg)I'_{ i , k}(t)\bigg]\\
=& \sum _ { i \in \mathcal{ S } } \sum _ { k \in \{ i \} \cup \mathcal{ C }}\mathbb { E } \bigg[ \bigg(  \overline { x } _ { i , k } - \tilde { x } _ { i , k } ( t ) \bigg)I'_{ i , k}(t) \mathds{1}_{\Delta X _ { i , k } ( t )}\bigg] \\\nonumber
\end{aligned}
\end{equation}
\begin{equation}\label{expectation:Z-3-t}
\begin{aligned}
+& \sum _ { i \in \mathcal{ S } } \sum _ { k \in \{ i \} \cup \mathcal{ C }}\mathbb { E } \bigg[ \bigg( \overline { x } _ { i , k } - \tilde { x } _ { i , k } ( t ) \bigg)I'_{ i , k}(t) \mathds{1}_{\Delta X _ { i , k } ^ {c}( t )}\bigg]\\
\overset{(a)}{\leq} & \sum _ { i \in \mathcal{ S } } \sum _ { k \in \{ i \} \cup \mathcal{ C }}\mathbb { E } \Bigg[ \underbrace{\bigg( \overline { x } _ { i , k } - \tilde { x } _ { i , k } ( t ) \bigg)I_{ i , k}(t) \mathds{1}_{\Delta X ^ {c}_ { i , k } ( t )}}_{K_1(t)}\Bigg], 
\end{aligned}
\end{equation}
where $(a)$ is due to $\overline { x } _ { i , k } - \tilde { x } _ { i , k } ( t ) \leq 0$ when the event occurs. 

Next, we define $K_1(t) \triangleq \Big(\overline { x } _ { i , k } - \tilde { x } _ { i , k } ( t ) \Big)I_{ i , k}(t) \mathds{1}_{\Delta X ^ {c}_ { i , k } ( t )}$. 
Then we consider the following two cases.

In the first case, \textit{i.e.}, when $t\leq t_{1}^{i,k}$, 
event $\Delta X^{c}_{i,k}(t)$ must not occur. This is because $\overline { x } _ { i , k } = 0$ due to $h_{i,k} = 0$ for $t\leq t_{1}^{i,k}$. Hence, for all $t\leq t_{1}^{i,k}$, we have 
\begin{equation}\label{expectation:K-1-t-case1}
\begin{aligned}
\mathbb{E} \bigg[K_1(t)\bigg] = 0.
\end{aligned}
\end{equation}

In the second case, when $t > t_{1}^{i,k}$, we suppose that event $\Delta X^{c}_{i,k}(t)$ occurs. 
Then we have $\overline { x } _ { i , k } - \tilde { x } _ { i , k } ( t ) \geq 0$ which, along with (\ref{equation: update ucb}), implies that
\begin{equation}
\Delta X _ { i , k } ( t ) = \bigg\{ \hat { x } _ { i , k } ( t - 1 ) - \overline x _ { i , k } > - \beta \sqrt { \frac {  \log t } { 4 h _ { i , k } ( t - 1 ) } } \bigg\}.
\end{equation}
Thus, we can upper bound $\mathbb{E} \Big[K_1(t)\Big]$ as follows.
\\
For all $t > t_{1}^{i,k}$, we have
\begin{equation}\label{expectation:K-1-t-case2}
\begin{aligned}
\vspace{-0.3em}
& \mathbb { E } \bigg[ K _ { 1 } ( t ) \bigg] \\
=\ & \mathbb { E } \bigg[ (\overline { x } _ { i , k }  - \tilde { x } _ { i , k } ( t ) )I ^ {' } _{i , k}(t)\mathds { 1 } _  { \Delta X _ { i, k}^{c} ( t ) }\bigg]\\
\stackrel{(a)}\leq\ & -x_{\text{min}}\mathbb { E } \bigg[\mathds { 1 } _  { \Delta X _ { i, k}^{c} ( t ) }\bigg]\\
=\ & -x_{\text{min}} \mathbb { P } \bigg\{ \hat { x } _ { i , k } ( t - 1 ) - \overline x _ { i , k } < - \beta \sqrt { \frac {  \log t } { 4 h _ { i , k } ( t - 1 ) } } \bigg\}\\
\stackrel{(b)}\leq& \frac{-x_{\text{min}}}{t^\frac{\beta^2}{2}},
\end{aligned}
\end{equation}
where inquality $(a)$ is due to $\overline { x } _ { i , k }  - \tilde { x } _ { i , k } ( t ) \leq -x_{\text{min}}$ and inequality $(b)$ is derived from Chernoff-Hoeffding bound.

By summing $\mathbb { E } \Big[ K _ { 1 } ( t ) \Big] $ over $t \in \{0,\ldots,T-1\}$ then applying (\ref{expectation:K-1-t-case1}) and (\ref{expectation:K-1-t-case2}), we have
\begin{equation}\label{expectation:K-1-t}
\begin{aligned}
\sum _ { t = 0 } ^ { T - 1 } \mathbb { E } \bigg[ K _ { 1 } ( t ) \bigg] & \leq -x_{\text{min}}\sum _ { t = t _ { 1 } ^ { i , k } + 1 } ^ { T - 1 } \frac { 1 } { t ^ { \frac{\beta^2}{2} } }\\
&\leq -x_{\text{min}}\sum _ { t = 1 } ^ { \infty } \frac { 1 } { t ^ { \frac{\beta^2}{2} } }\\
&= -G_{\beta}x_{\text{min}}.
\end{aligned}
\end{equation}
Likewise, we have
\begin{equation}\label{expectation:Z-3-t}
\begin{aligned}
\sum _ { t = 0 } ^ { T - 1 } \mathbb { E } \bigg[ Z _ { 3 } ( t ) \bigg] \leq -|S|(|C|+1)G_{\beta}x_{\text{min}}.
\end{aligned}
\end{equation}
According to (\ref{inequation: upper_bound_of_average_regret}), 
(\ref{definition: x_min}), and 
\begin{equation}
	Z_1(t) \leq V(Z_2(t) + Z_3(t)),	
\end{equation}
we have
\begin{equation}
\begin{aligned}
\nonumber
&\frac { 1 } { T } \sum _ { t = 0 } ^ { T - 1 } \mathbb { E } [ \Delta R ( t ) ]  \leq \frac { B } { V \lambda _ { \text{max} } } + \frac { 1 } { T V } \sum _ { t = 0 } ^ { T - 1 } \mathbb { E } \bigg[ Z _ { 1 } ( t ) \bigg]\\
=& \frac { B } { V \lambda _ { \text{max }} } + \frac { 1 } { T V }V(\sum _ { t = 0 } ^ { T - 1 } \mathbb { E } \bigg[ Z _ { 2 } ( t ) \bigg] + \sum _ { t = 0 } ^ { T - 1 } \mathbb { E } \bigg[ Z _ { 3 } ( t ) \bigg])\\
\leq & \frac { B } { V \lambda _ { \text{max} } } + \frac { 1 } { T }\bigg[(| S | ( | C | + 1 ) \bigg( 2 \beta \sqrt { T \log T } - \\
& ( G_{\beta} + 1 ) x _ { \text{min} } \Big) \Big] - \frac { 1 } { T } |S|(|C|+1)G_{\beta}x_{\text{min}}\\
\end{aligned}
\end{equation}
\begin{equation}
\begin{aligned}
=& \frac { B } { V \lambda _ { \text{max} } } + \frac { | S | ( | C | + 1 ) } { T }\bigg( 2 \beta \sqrt { T \log T } - ( 2G_{\beta} + 1 ) x _ { \text{min} } \bigg)\\
=& \frac { B } { V \lambda _ { \text{max} } } + \frac { | S | ( | C | + 1 ) } { T }[ 2 \beta \sqrt { T \log T }+ \\
& ( 2G_{\beta} + 1 )\max\{w_{\text{max}},m_{\text{max}}\} ],
\end{aligned}
\end{equation}
where constant $B$ is defined in (\ref{definition: B}) 
and $G_{\beta} \triangleq \sum _ { t = 1 } ^ { \infty } t ^ { -\frac { \beta ^ { 2 } } { 2 } } $.\\
\IEEEQED

\end{document}